\documentclass{jfm}
\usepackage{graphicx}
\usepackage{epstopdf, epsfig}
\usepackage{ar}
\usepackage{caption}
\usepackage{subcaption}
\usepackage{amsmath}
\usepackage{amsfonts}
\usepackage{amssymb}
\usepackage{multirow}
\usepackage{xcolor}
\usepackage{comment}
\usepackage{float}

\usepackage{booktabs}
\usepackage{siunitx}
\usepackage{tabularray}
\UseTblrLibrary{booktabs,siunitx}

\def\vK{von K\'{a}rm\'{a}n }

\shorttitle{Vortex capture dictates efficiency in three-hydrofoil schools}
\shortauthor{P. C. Ormonde, Y. Zhu, D. Quinn, K. Moored}

\title{Rather than drafting, vortex capture dictates efficiency in three-hydrofoil schools}

\author{Pedro C. Ormonde\aff{1,3}
  \corresp{\email{pco@brown.edu}},
  Yuanhang Zhu\aff{2,4},
  Daniel Quinn\aff{2},
  \and Keith W. Moored\aff{1}}

\affiliation{\aff{1}Department of Mechanical Engineering and Mechanics, Lehigh University, Bethlehem, PA 18015, USA
\aff{2}Department of Mechanical and Aerospace Engineering, University of Virginia, Charlottesville, VA 22094, USA
\aff{3}Center for Fluid Mechanics, School of Engineering, Brown University, Hope St., Providence, RI, 02912, USA
\aff{4}Department of Mechanical Engineering, University of California, Riverside, CA, 92521, USA}

\begin{document}

\maketitle

\begin{abstract}
Three-dimensional experiments are presented on a school of three pitching hydrofoils. Two side-by-side leader foils maintain the same relative positions while the location of a third follower foil is varied. Force and flow measurements detail the mechanisms that drive the school to achieve collective thrust and efficiency that are 58\% and 24\% higher than isolated foils, respectively. Traditional drafting involves positioning yourself in the wake of an upstream object. In wakes with a net momentum deficit, drafting reduces drag by lowering oncoming flow speed. By contrast, wakes from oscillatory swimmers feature strong momentum surplus regions, which increases drag by increasing the oncoming flow. Despite that, our results show that the best performance benefits occur for compact schools where the follower is directly in the vortex wake of a leader, whereas regions of reduced mean flow do not improve performance. The thrust and efficiency benefits are shown to be driven by vortex-body interactions that increase the thrust and efficiency of the follower and by body-to-body upstream interactions that reduce the power of the leaders. There is an optimal spatial phase to maximize the thrust and efficiency of the follower that depends upon the actual wake wavelength rather than the estimated wavelength used in previous literature. Moreover, wake breakdown, and its associated elimination of vortex-body performance benefits, is not observed within at least three chord lengths downstream of the leaders. Lastly, measurements of the cross-stream stability of the downstream foil indicate that compact, high-performance formations may require active control strategies in order to maintain their organization and maximise the hydrodynamic benefits of schooling.

\end{abstract}



\section{Introduction} \label{sec:intro}

The high efficiency and manoeuvrability of aquatic animals \citep{fish2020bio} have driven the design of numerous underwater vehicles using bio-inspired propulsive mechanisms \citep{lauder2011bioinspiration,roper2011review,siddall2014launching,katzschmann2018exploration,zhu2019tuna,zhong2021tunable,white2021tunabot}. Now, attention is turning to operating these bio-inspired vehicles as multi-agent collectives or schools \cite[]{berlinger2021implicit} to allow them to perform distributed tasks and to harness the hydrodynamic benefits of schooling, such as the reduced energy expenditure that fish in a school experience \cite[]{zhang2024energy}. Previous hydrodynamic studies on pairs of swimmers have found that canonical formations such as in-line formations \citep{Boschitsch2014,ramananarivo2016flow,muscutt2017performance,kurt2018flow,Heydari2020,kurt2021high} and side-by-side formations \citep{verma2018efficient,Kurt2020,zhong2021aspect,ormonde2021two,han2023revealing} have the potential to substantially increase the thrust and/or efficiency when there is proper synchronisation between swimmers. 

Several mechanisms have been proposed to be behind these schooling benefits. \cite{weihs1973hydromechanics,weihs1975some} proposed an ideal model of schooling where fish would minimise their energy expenditure by arranging themselves into two-dimensional layers with every other row of fish staggered, giving rise to the now classical diamond formation used as a model arrangement for several studies \citep{stocker1999models,hemelrijk2015increased,pan2022effects,kelly2023hydrodynamics,wei2023hydrodynamic,kelly2024effects}. Weihs argued that this formation would exploit several mechanisms at once leading to reduced energy expenditure.  One benefit would occur when fish in  each row would ideally synchronise with an out-of-phase relationship to their nearest neighbour. This would lead to direct body-body interactions akin to ground effect that would increase their thrust \citep{Dong2007,Quinn2014,Dewey2014,zhong2021aspect}.  A second benefit would be derived when a row of fish downstream of a leading row would swim in between the reverse \vK vortex streets shed by the leaders. This would cause the follower fish to experience a reduced flow region in the \textit{time-average} and, consequently, reduced drag -- a form of a drafting mechanism. Later, it was appreciated that \textit{unsteady} vortex-body interactions could also lead to substantial thrust and efficiency enhancements for a follower when its tail motion is synchronised with the induced \textit{time-varying} flow from an impinging or nearby vortex street \citep{Gopalkrishnan1994,streitlien1996efficient,liao2003karman,liao2004neuromuscular,beal2006passive,Akhtar2007,Boschitsch2014, portugal2014upwash,kurt2018flow,li2020vortex,alben2021collective,han2022propulsive} thereby maximising the follower's effective angle of attack \citep{muscutt2017performance}. Contrary to these two-dimensional studies of a pair of swimmers, studies examining three-dimensional swimmers in \textit{infinite} schools show that wake breakdown eliminates coherent vortex-body interactions and leaves only direct body-body interactions to provide a schooling benefit \citep{Daghooghi2015}. Yet, three-dimensional fish-like swimmers can experience performance boosting vortex-body interactions when swimming in a \textit{pair} \citep{verma2018efficient} or not \citep{li2019energetics} depending upon their kinematics and/or geometry.  Understanding the interplay and applicability of schooling mechanisms from body-body interactions and drafting to vortex-body interactions and wake breakdown for three-dimensional swimmers in school sizes beyond a pair of swimmers has yet to be fully understood.

Beyond the direct performance benefits that occur from various schooling mechanisms it is also known that these interactions can give rise to forces that push and pull swimmers to settle into hydrodynamically stable formations, a phenomenon known as the Lighthill conjecture \citep{Ramananarivo2016}. Previous work on unconstrained foils has discovered streamwise stable \citep{becker2015hydrodynamic,peng2018collective,newbolt2019flow,heydari2021school,wei2023hydrodynamic}, and both streamwise and cross-stream stable formations of two-dimensional \citep{lin2021flow,lin2022two} and three-dimensional \citep{ormonde2021two} swimmers. Constrained or tethered swimmers have also been used to approximate free-swimming states and to characterise the stability of formations in order to qualify their performance benefits \citep{li2019energetics}. To date, our knowledge of the stability of formations as the school size increases beyond a pair of swimmers is quite limited to only a few studies \citep{wei2023hydrodynamic,newbolt2024flow}.

Here, we present new experiments on a school of three pitching hydrofoils and advance our understanding of schooling interactions in several ways. This study, to the authors' knowledge, is the first \citep{ligman2023comprehensive} to experimentally detail schooling interactions among more than two, three-dimensional swimmers at moderately-high ($Re>5$,$000$), biologically-relevant Reynolds numbers. Inspired by Weihs' schooling hypothesis \citep{weihs1973hydromechanics,weihs1975some} we study a school of two leaders and one follower where we can directly examine the ``back-half" of the classic diamond formation and probe the associated schooling mechanisms of wake breakdown, drafting, vortex-body interactions, and body-body interactions. We also further examine the stability characteristics of the three-foil school. The paper is organised as follows. Section~\ref{sec:methods} describes the experimental setup and methodology used throughout the study. Section~\ref{sec:flowfields} details the three-dimensional flowfields generated by the small school. Section~\ref{sec:FollowerPerf} probe drafting, vortex-body, and body-body mechanisms for the downstream foil. Section~\ref{sec:LeaderPerf} presents the performance of the leaders. Section~\ref{sec:3foilSchool} presents the collective performance of the school and examines its stability. Section~\ref{sec:conclusions} provides a discussion of the results and summarises the conclusions of the study.

\section{Methods}
\label{sec:methods}
To study the flow interactions, and their implications, that occur in fish schools we focus on an idealized model of a fish school as three pitching hydrofoils, as adopted previously \cite[]{moored2019inviscid}. Pitching hydrofoils capture the salient features of unsteady flows of circulatory forces, added mass forces, and vortex shedding that are key characteristics underlying the hydrodynamics of fish swimming. These hydrofoils can be viewed as a representation of the caudal fins of fish, which, for many species, are the primary propulsive elements responsible for generating the bulk of a swimmer's thrust~\citep{smits2019undulatory}. 

\subsection{Water-channel experiments}
\begin{figure}
  \centering
  \includegraphics[width=0.8\textwidth]{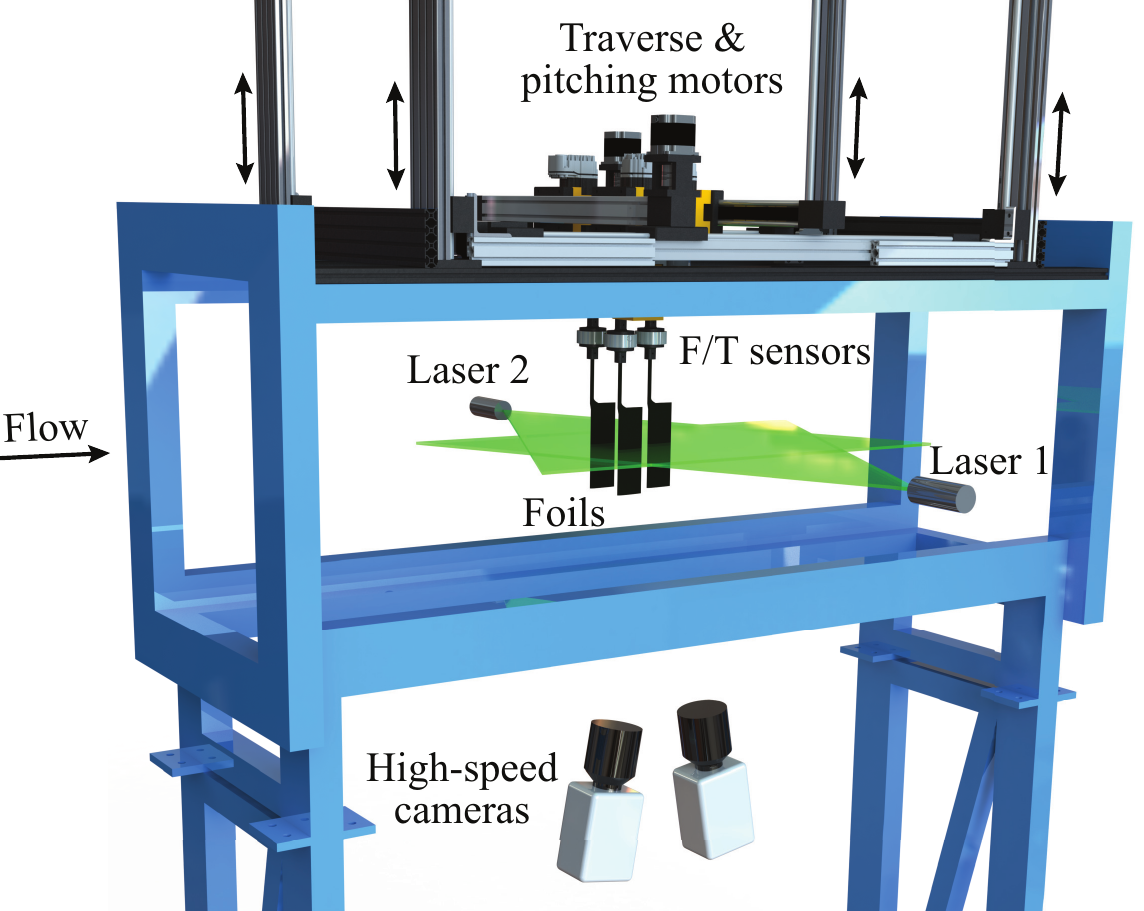}
  \caption{Water channel experimental setup. The follower positioning in the horizontal $x-y$ plane is automated using a traverse system. Stereo PIV is performed at a fixed horizontal plane and the multiple spanwise PIV layers are obtained by traversing all three foils up and down using a vertical traverse system.}
\label{fig:UVAsetup}
\end{figure}

The three pitching hydrofoils are machined-aluminium with a NACA 0012 cross-section, a chord length of $c= \ $0.0488 m, and an aspect ratio of $\AR= 3$. The hydrofoils are immersed in a recirculating water channel ($W \times H \times L= 0.38 \times 0.45 \times 1.52$ m) at the University of Virginia. Figure~\ref{fig:UVAsetup} shows a schematic of the water channel setup.  The flow speed is maintained at $U=0.174$ m/s using an ultrasonic flow meter (Dynasonics Series TFXB). The foils are constrained at fixed positions during each trial. The foil shaft is connected to a six-axis force/torque transducer (ATI Mini40) to measure the hydrodynamic loads experienced by the foil, and an optical encoder (US Digital E2-5000) measures the pitching angle. The foil is driven by a servo motor (Teknic CPM-MCPV-2341S-ELN) coupled to a 5:1 gearbox (SureGear PGCN23-0525). The positioning of the three foils is enabled by a four-axis motorised traverse system, in which the follower is traversed in the $x$- and $y$-directions independently. A sinusoidal oscillatory pitching motion about the leading-edge axis is prescribed to all three swimmers. 

We use multi-layer phase-averaged stereoscopic particle image velocimetry (sPIV) to measure the three-dimensional flow field around the pitching foils. Similar techniques have been used previously~\citep[]{king2018experimental,zhong2021aspect,zhu2023flow,zhu2025wavenumber}. We seed the water flow using neutrally buoyant 50 $\mu$m silver-coated hollow ceramic particles (Potters Industries), and illuminate the particles using two 5 mm-thickness laser sheets firing from each side of the water channel to minimize foil shadows. The laser sheets are created using continuous wave lasers (532 nm, 5W Raypower MGL-W-532 and 10W CNI MGL-W-532A) with laser guiding arms and sheet optics. Two high-speed cameras (Phantom SpeedSense M341, $2956\times1877$ pixels) with Scheimpflug adaptors (Dantec Dynamics) and 50 mm lenses (Zeiss) beneath the water channel are used for recording the PIV image pairs. The raw images are processed using Dantec Dynamic Studio 6.9 by an adaptive PIV algorithm (minimum interrogation window $32\times32$ pixels, maximum interrogation window $64\times64$ pixels). To obtain the 3D velocity field, we fix the laser sheet position and traverse the foils in the $z$-axis with a step size of 5 mm. For each spanwise layer, we phase-average 750 instantaneous 2D3C velocity fields over 25 cycles (i.e. 30 vector fields per cycle). Due to symmetry, we measure the flow field of the bottom half of the foils plus 2 cm below the foil tip (20 layers) and then mirror the flow field about the mid-span plane to construct the 3D3C velocity field (39 layers). The 3D velocity field is reconstructed using MATLAB (R2023a). 

To visualise the three-dimensional flow structures, we use the $Q$-criterion \citep{jeong1995identification}, $Q=0.5(\Vert\boldsymbol{\Omega}\Vert^2-\Vert\boldsymbol{\mathrm{S}}\Vert^2)$, where $Q$ is the second invariant of the velocity gradient tensor, $\boldsymbol{\Omega}$ is the vorticity tensor and $\boldsymbol{\mathrm{S}}$ is the strain-rate tensor. The flow structures are plotted as iso-$Q=10$ surfaces coloured by the non-dimensional spanwise vorticity $\omega_zc/U$. 

\begin{figure}
  \centering
     \begin{subfigure}[b]{0.45\textwidth}
         \centering
         \includegraphics[width=\textwidth]{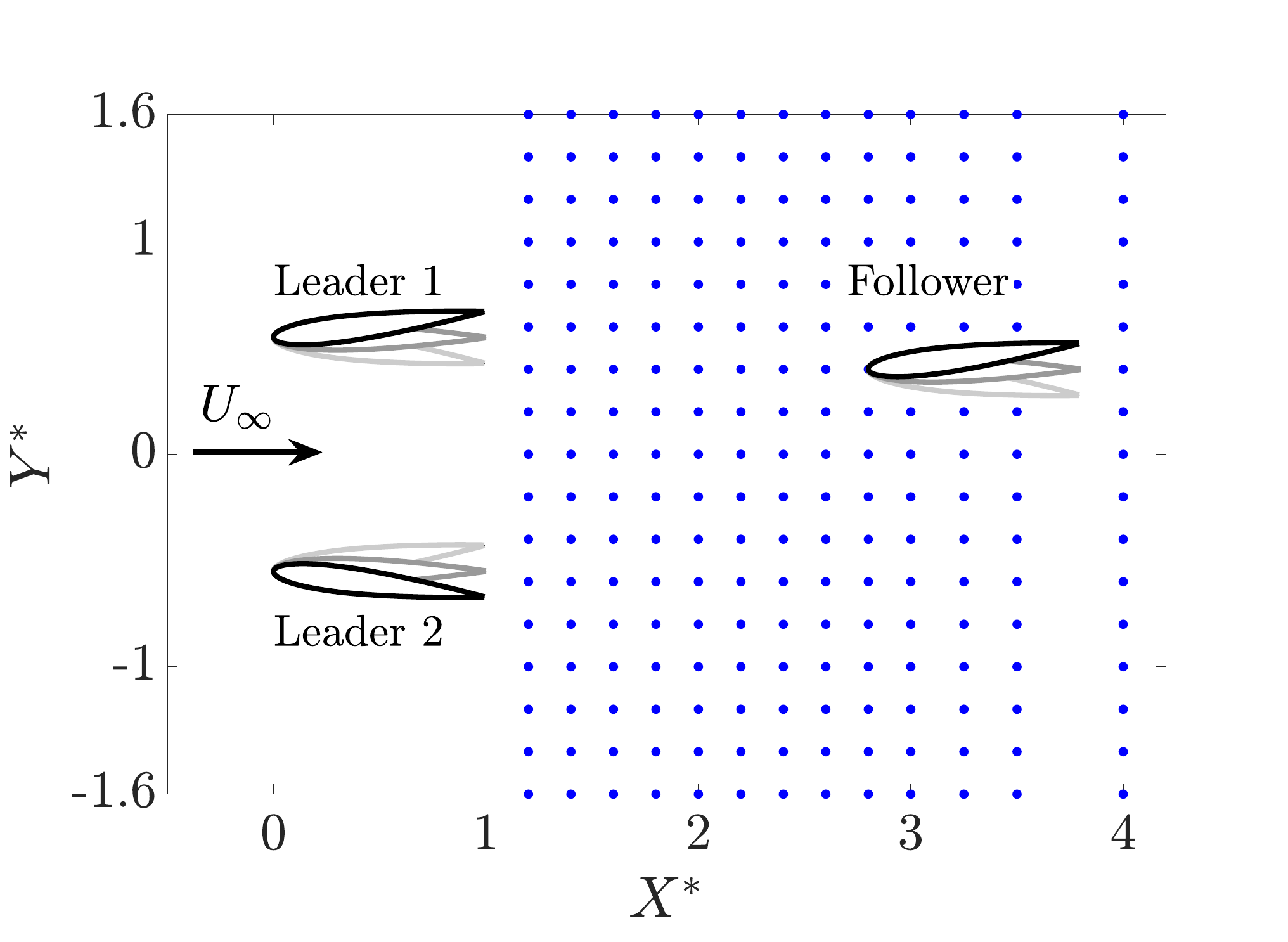}
         \caption{In-phase leaders}
     \end{subfigure}
     \hspace{0cm}
     \begin{subfigure}[b]{0.45\textwidth}
         \centering
         \includegraphics[width=\textwidth]{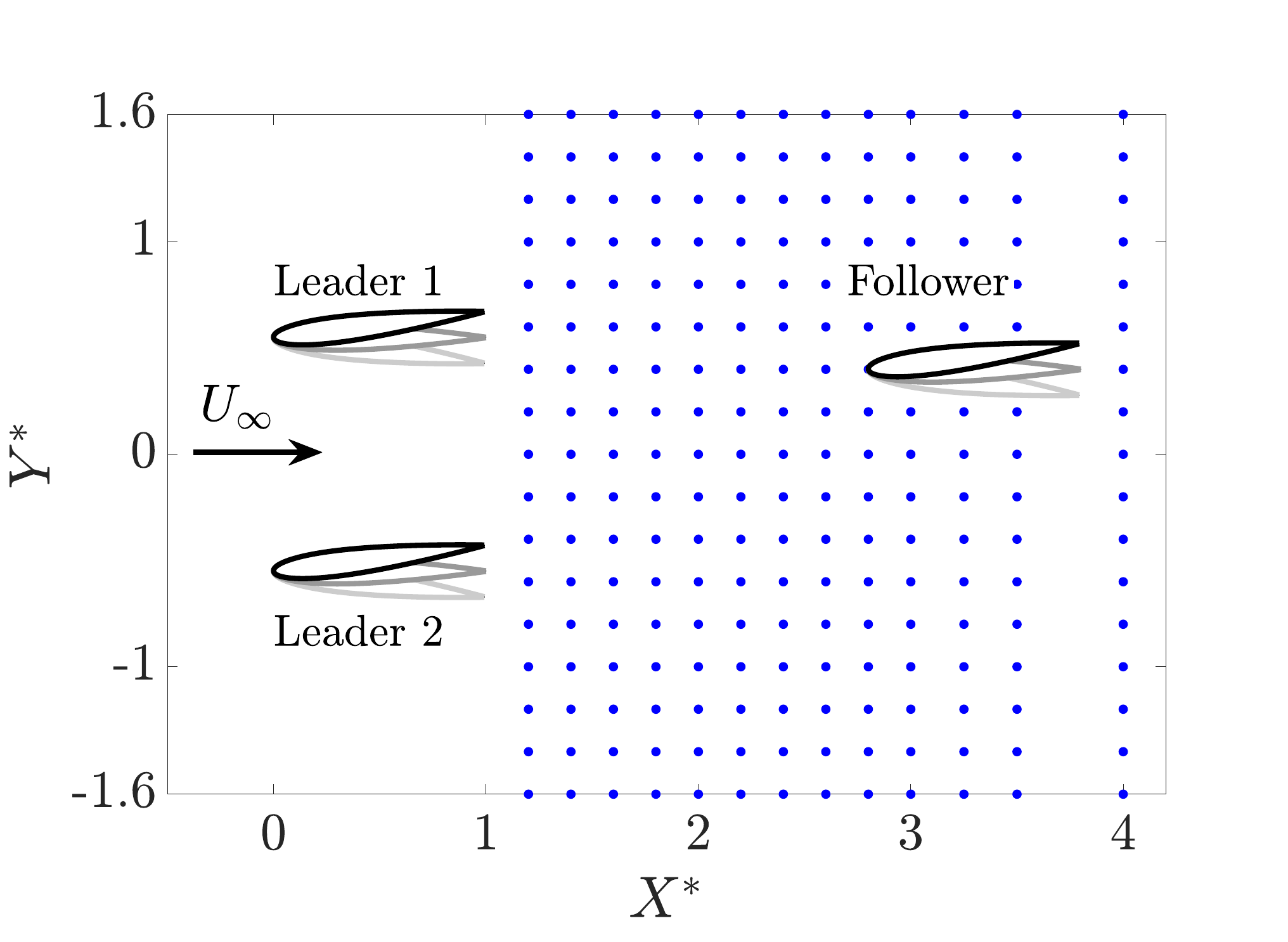}
         \caption{Out-of-phase leaders}
     \end{subfigure}
     \caption{Schematics of the arrangement of the hydrofoils in the experimental domain. Dots represent the locations of the follower's leading edge. The domain is $1.2\leq X^* \leq 4.0$, $-1.6 \leq Y^* \leq 1.6$. Leader 1 and the follower are always in-phase, and leader 2 is either in-phase with leader 1 ($\phi = 0$), or out-of-phase ($\phi = \pi$).}
\label{fig:domain}
\end{figure}

One foil, designated \emph{leader 1}, is used as the reference with kinematics $\theta_{1}(t) = \theta_0 \sin(2\pi f t)$ where $\theta_{1}(t)$ is the instantaneous pitching angle, $\theta_0$ is the pitching amplitude, and $f$ is the oscillation frequency. The second foil, denoted as \emph{leader 2}, is positioned to the side of leader 1 with prescribed kinematics $\theta_{2}(t) = \theta_0 \sin(2\upi f t + \phi)$, where $\phi$ is the synchrony, or phase difference of leader 2 relative to leader 1. Experiments were performed for the leaders swimming in-phase ($\phi = 0$) and out-of-phase ($\phi = \upi$). The third foil, designated as the \emph{follower}, oscillates in-phase with leader 1 with identical kinematics, i.e., $\theta_3(t) = \theta_1(t)$. All three foils have a dimensionless peak-to-peak amplitude of motion of $A^* = A/c = 0.24$ with $A=\sin{(2\theta_0)}$, Strouhal number $St = f A/U = 0.29$, reduced frequency $k = f c/U = 1.14$, and chord-based Reynolds number $Re = $ 8,450. The relative positions of the leaders remain constant throughout the entire study in a side-by-side formation with a dimensionless cross-stream distance of $D_{2,1}^* = D_{2,1}/c=1.1$, which is the closest spacing possible for the experimental apparatus. The position of the follower is varied throughout a two-dimensional grid presented in Figure \ref{fig:domain}, ranging from $1.2 \leq X^* \leq 4.0$, $-1.6 \leq Y^* \leq 1.6$. A constant cross-stream grid spacing of $\Delta Y^*=0.2c$ is employed. The streamwise spacing $\Delta X^* = 0.2$c is used in the region $1.2 \leq X^* \leq 3.0$. Further downstream data is collected at points $X^*=[3.25, \ 3.5, \ 4.0]$. Each blue dot represents a position of the leading edge of the follower where measurements were collected.

Measurements of the thrust ($T$), lift ($L$) and pitching moment ($M_z$), allows us to calculate the thrust ($C_T$), lift ($C_L$) and power ($C_P$) coefficients as well as the efficiency ($\eta$) for each foil
\begin{equation} 
C_T = \frac{\overline{T}}{q_\text{dyn} c s } \text{,} \quad C_L = \frac{\overline{L}}{q_\text{dyn} c s } \text{,} \quad C_P = \frac{\overline{P}}{q_\text{dyn} U_{\infty} c s } \text{,} \quad \eta = \frac{C_T}{C_P}
\end{equation} 

\noindent where $q_\text{dyn} = \frac{1}{2} \rho U^2$ is the dynamic pressure, $\rho$ is the fluid density, $s$ is the foil span and $U$ is the freestream speed. These four quantities can also be defined for the collective denoted with a $C$ subscript: 
\begin{equation}
         C_{T,C} = \frac{ {\overline{T}}_{1} + {\overline{T}}_{2} + {\overline{T}}_{3} }{ q_\text{dyn} (3 c s) } \text{,} \quad  C_{L,C} = \frac{ {\overline{L}}_{1} + {\overline{L}}_{2} + {\overline{L}}_{3} }{ q_\text{dyn} (3 c s) } \text{,} \quad  C_{P,C} = \frac{ {\overline{P}}_{1} + {\overline{P}}_{2} + {\overline{P}}_{3} }{ q_\text{dyn} U (3 c s) } \text{,} \quad  \eta_{C} = \frac{C_{T,C}}{C_{P,C}} 
\end{equation}

In this study we compare the performance of a foil that is in a school to that of a single, isolated foil. The forces and power coefficients are thus normalised by their equivalent isolated foil quantities, noted with the $(\cdot)^{\text{iso}}$ superscript. The normalised quantities are noted with the $(\cdot)^*$ superscript and are defined as
\begin{equation}
    C^*_T = \frac{C_T}{C_{T}^\text{iso}} \text{,} \quad C^*_L = \frac{C_L}{C_{L}^\text{iso}} \text{,} \quad C^*_P = \frac{C_P}{C_{P}^\text{iso}} \text{,} \quad \eta^* = \frac{\eta}{\eta^{\text{iso}}}
\end{equation}

The measured thrust of the isolated foils are $C_{T,1}^\text{iso} = 0.18 \pm 0.03 $, $C_{T,2}^\text{iso} = 0.17 \pm 0.02 $ and $C_{T,3}^\text{iso} = 0.12 \pm 0.01 $. The values for the static drag coefficients are $C_{D,1}^\text{static} = 0.16 \pm 0.02$, $C_{D,2}^\text{static} = 0.12 \pm 0.02$ and $C_{D,3}^\text{static} = 0.12 \pm 0.01$. The power coefficients are $C_{P,1}^\text{iso} = 0.989\pm 0.006$, $C_{P,2}^\text{iso} = 0.836 \pm 0.001$ and $C_{P,3}^\text{iso} = 0.739\pm 0.001$, and the efficiencies are $\eta_1^{\text{iso}} = 0.18\pm 0.03$, $\eta_2^{\text{iso}} = 0.21\pm 0.02$ and $\eta_3^{\text{iso}} = 0.17\pm 0.01$.

The average performance metrics for the three-foil school can also be compared to the collective performance of three isolated foils performance to define normalized \textit{collective} performance metrics as
\begin{equation}
    C^*_{T,C} = \frac{C_{T,C}}{C_{T,C}^\text{iso}} \text{,} \quad C^*_{L,C} = \frac{C_{L,C}}{C_{L,C}^\text{iso}} \text{,} \quad C^*_{P,C} = \frac{C_{P,C}}{C_{P,C}^\text{iso}} \text{,} \quad \eta_C^* = \frac{\eta_C}{\eta_{C}^\text{iso}}.
\end{equation}

\section{Three-dimensional flow fields}
\label{sec:flowfields}

\begin{figure}
    \centering
    \includegraphics[width=\linewidth]{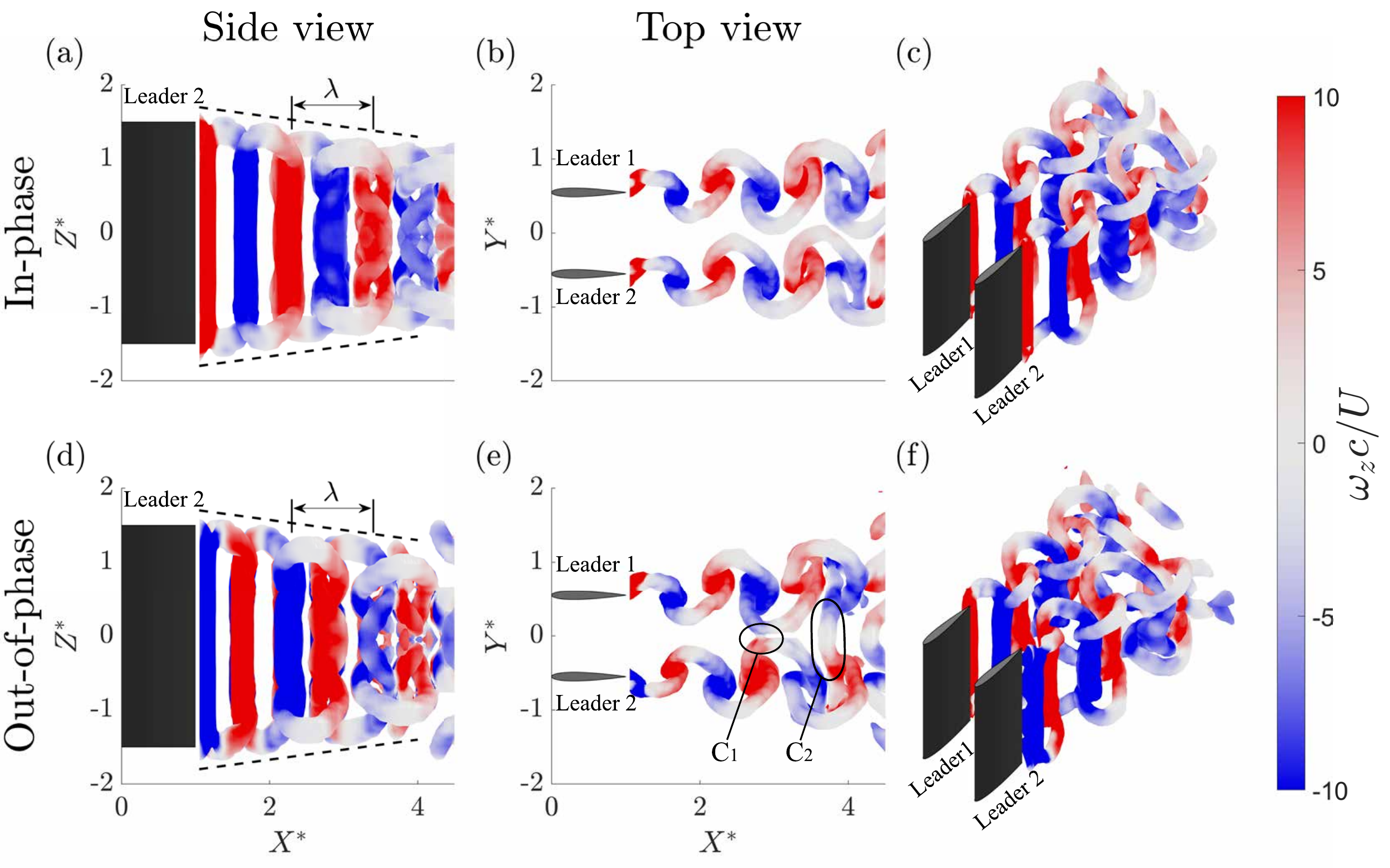}
    \caption{Isosurfaces of $Q=10$ from phase-averaged PIV data of the two leaders in the absence of the follower. (a) and (d): Side view. Black dashed lines highlight the spanwise compression of the wake structures and their wake wavelength $\lambda$ is marked. (b) and (e): Top view. For the out-of-phase case (e) the iso-surfaces from the two wakes merge together, starting around $X^*=3.8$, as the two horizontally-oriented portions of the surfaces (light grey color) generated by the wing tips make contact in region $C_1$, eventually merging completely as highlighted by $C_2$. (c) and (f): Perspective view. }
    \label{fig:WakeLeaders_3D}
\end{figure}

We start by presenting the three-dimensional flow fields generated by the two side-by-side leaders in the absence of the downstream follower. The leaders are positioned side by side with leader 1 located at $(X^*_1,\,Y^*_1) = (0,\,0.55)$ and leader 2 at $(X^*_2,\,Y^*_2) = (0,\,-0.55)$. Figure \ref{fig:WakeLeaders_3D} presents vortex structures that form the wake shed by the side-by-side leaders. All plots shown are constructed using either time-averaged or phase-averaged PIV data for a zero pitching angle ($\theta = 0$), which is at the beginning of the pitching cycle $t^* = t/T = 0$ ($T = 1/f$ is the cycle period). For the in-phase leaders (Figure \ref{fig:WakeLeaders_3D}a-c), both foils are pitching counter-clockwise ($-\hat{z}$), and for the out-of-phase leaders, leader 1 is pitching counter-clockwise while leader 2 is pitching clockwise ($+\hat{z}$). The three-dimensional structure of the wake shows the shedding of alternating vortex rings that are intertwined amongst themselves, generated from the tips and trailing edge of the foils once every half-cycle of the pitching motion \citep{buchholz2008wake,king2018experimental}. The wake structures narrow along the spanwise dimension $z$ as they advect downstream, a phenomenon observed for low aspect-ratio rectangular panels \citep{buchholz2008wake} and trapezoidal panels of aspect ratio 4.17 \citep{king2018experimental}. The dashed lines on Figures \ref{fig:WakeLeaders_3D}a and \ref{fig:WakeLeaders_3D}d highlight this spanwise compression.    
\begin{figure}
    \centering
    \includegraphics[width=\linewidth]{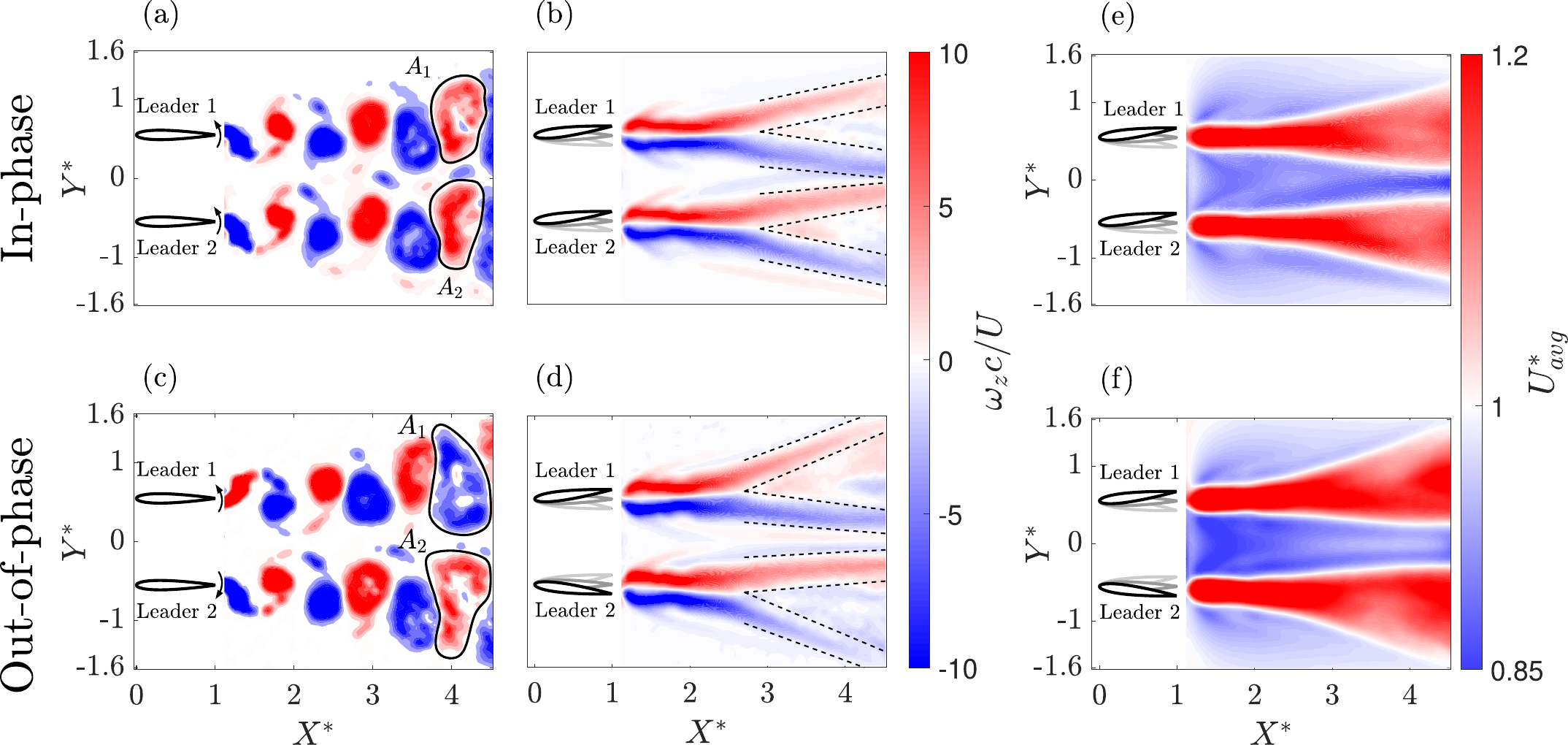}
    \caption{Flowfield at the midspan plane $z=0$ for two leaders. (a) and (b): Cycle-average vorticity for the in-phase (a) and out-of-phase (c) cases. $A_1$ and $A_2$ highlights regions of more pronounced breakdown of the vortices shed by Leader 1 and Leader 2, respectively. Contours presented for beginning of pitching cycle $t/T = 0$. (b) and (d): Time-average vorticity contours for in-phase (b) and out-of-phase (d) cases. Dashed lines highlight the deflection of the positive (red) and negative (blue) vortices due to the interaction between the two wakes. (e) and (f): Mean streamwise velocity fields for in-phase (e) and out-of-phase (f) cases. The dark blue region between the out-of-phase leaders in (f) show a reduction of up to 17\% in the mean streamwise speed due to the synchrony between the wakes for $1.1\leq X^* \leq 2.2$.}
    \label{fig:Leaders_2Dwakes}
\end{figure}

Figures~\ref{fig:Leaders_2Dwakes}a and \ref{fig:Leaders_2Dwakes}b show the $z$-component of vorticity, $\omega_z$, at the midspan of the foils ($z = 0$) and at $t^* = 0$. A typical reverse \vK vortex street is observed up to 2.5 wake wavelengths downstream of the trailing edge (at about $X^*=4$). This can be observed at any given spanwise location within $-1<Z^*<1$ from the 3D structures shown in Figure~\ref{fig:WakeLeaders_3D}, with well defined regions of spanwise vorticity of alternating sign. The wakes thus retain features characteristic of 2D flows for a relatively long convective time due to the aspect ratio ($AR=3$) and Strouhal number ($St=0.29$) chosen for this study. Further downstream, at $X^*>4$, the wakes start breaking down and the spanwise oriented sections of the vortex rings from both hydrofoils unravel, interact, and eventually merge, leading to a disruption of the vortical structures observed upstream. The breakdown of coherent vortex structures is even more pronounced for the out-of-phase case, as can be seen in Figures \ref{fig:WakeLeaders_3D}d-f. Figure \ref{fig:WakeLeaders_3D}e shows the coherent vortex structures from both leaders making contact as highlighted by region $C_1$, and completely merging downstream in region $C_2$, which enhances their breakdown. The mean vorticity contours depicted in Figures~\ref{fig:Leaders_2Dwakes}b and \ref{fig:Leaders_2Dwakes}d also aid in visualising how the interaction of the two wakes alters their development and advection trajectories. For the in-phase case the clockwise vortices (blue) from leader 1 and counter-clockwise vortices (red) from leader 2 curve towards each other starting at $X^*\approx 2.8$,  eventually meeting at the symmetry line $y=0$ at $X^*=4.5$, spreading out the wakes of each one of the leaders along the cross-stream dimension. Figures~\ref{fig:Leaders_2Dwakes}e and \ref{fig:Leaders_2Dwakes}f present the time-average, streamwise velocity $\overline{u}^*$ in the wake of the leaders at the midspan ($z=0$). The flowfields show two distinct jets (coloured red) with mean flow speeds up to 20\% higher than the freestream. The region between the two jets (blue) shows a reduction of the streamwise flow speed, with maximum reductions of 11\% and 17\% for the in-phase and out-of-phase cases, respectively. According to the \emph{drafting mechanism} proposed by \cite{weihs1973hydromechanics}, a third swimmer (follower) positioned within this region would benefit from a reduction in the viscous drag force, thus improving its thrust and efficiency. In Section~\ref{subsec:3foil_drafting} it is shown that the performance of the follower \textit{decreases} in this region, contrary to the predictions of the drafting mechanism. In fact, the best thrust benefits for the follower occur in the \textit{accelerated} flow region of the jets where there are direct vortex-body interactions that are regulated by the phase synchrony between the pitching motion of the follower and the impinging wake vortices -- something not considered in the drafting hypothesis proposed in \cite{weihs1973hydromechanics}. It will be shown that the hypothesised performance gains originating from viscous drag reduction in the reduced flow region are not observed for a school of three pitching hydrofoils. 

\begin{figure}
    \centering
    \includegraphics[width=\linewidth]{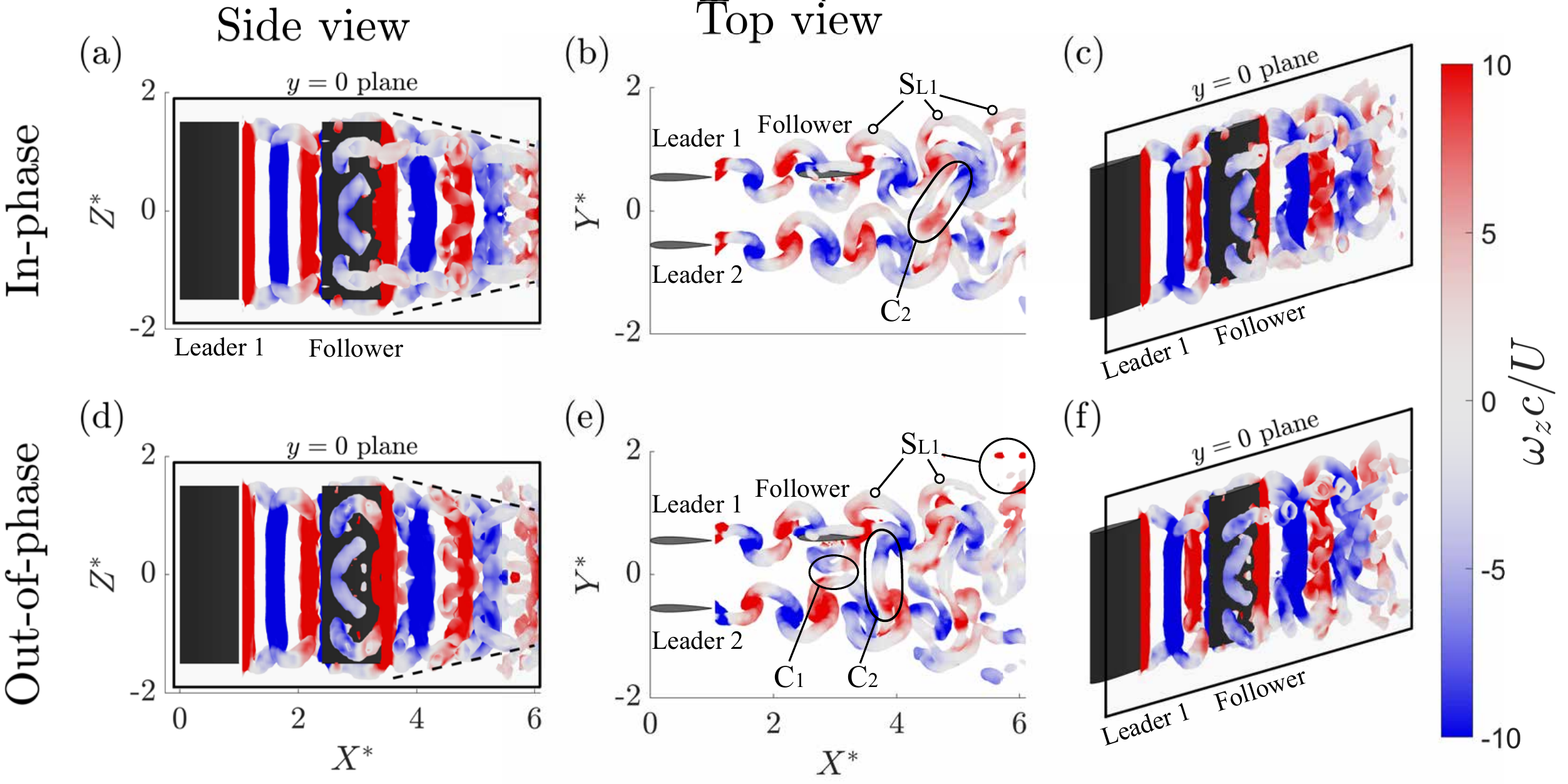}
    \caption{Isosurfaces of $Q=10$ from phase-averaged PIV data of the 3-foil school. (a) and (d): Side view of the flowfield structures located behind the vertical plane $y=0$. The wake of Leader 1 impinges directly onto the Follower. Black dashed lines show the spanwise compression of the wake structures generated by the Follower. (b) and (e): Top view. The structures highlighted by $S_{L1}$ are generated by Leader 1 and greatly deformed by the presence of the Follower for both the in-phase (b) and out-of-phase (e) cases. The iso-surfaces from the two wakes merge, starting around $X^*=4$ for the in-phase case (b) and $X^*=3$ for the out-of-phase case (e), as the two horizontally-oriented portions of the surfaces (light grey color) generated by the wing tips make contact in region $C_1$, eventually merging completely as highlighted by $C_2$. (c) and (f): Perspective view. Only structures behind the vertical plane $y=0$ are depicted for clarity.}
    \label{fig:3foils_3Dwake}
\end{figure}

\begin{figure}
    \centering
    \includegraphics[width=\linewidth]{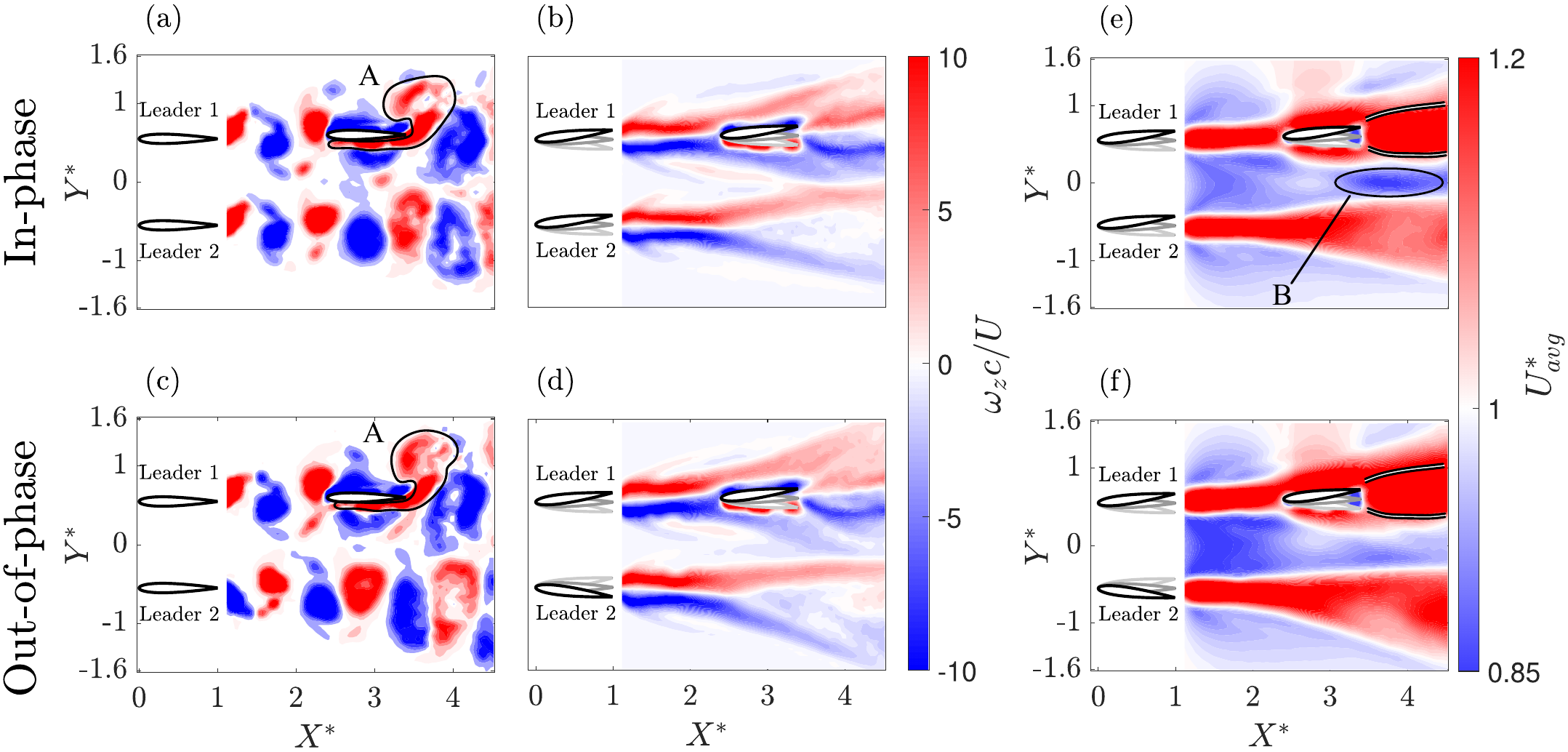}
    \caption{Flowfield at the midspan plane $z=0$ for the minimal school. (a) and (c): Cycle-average vorticity for the in-phase (a) and out-of-phase (c) cases. Region $A$ highlights the positive (red) vorticity generated at the left-side surface of the Follower merging to the wake vortex shed by Leader 1. Contours presented for beginning of pitching cycle $t/T = 0$. (b) and (d): Time-average vorticity contours for in-phase (b) and out-of-phase (d) cases. (e) and (f): Mean streamwise velocity fields for in-phase (e) and out-of-phase (f) cases. Black and white lines downstream of the Follower highlight the Follower mean streamwise jets of increased flow speed. Region B in (e) highlights an area of reduced flow speed along the symmetry plane $y=0$ for the in-phase case.}
    \label{fig:3foils_2Dwake}
\end{figure}

Figure \ref{fig:3foils_3Dwake} shows the vortex structures measured for the three foil school. Here the follower is located almost directly behind leader 1 at $(X^*,\,Y^*) = (2.4,\,0.6)$ for both the in-phase and out-of-phase cases. Figures \ref{fig:3foils_3Dwake}a, \ref{fig:3foils_3Dwake}c, \ref{fig:3foils_3Dwake}d and \ref{fig:3foils_3Dwake}f show the flow structures behind the vertical plane $y=0$, i.e., only the structures at $Y^*>0$ are displayed and the structures in front of the plane ($Y^*<0$) are omitted for clarity. Figures \ref{fig:3foils_3Dwake}a and \ref{fig:3foils_3Dwake}d show leader 1 and the follower from a side view, as seen from their left-hand side. The wake structures from leader 1 can be seen impinging onto the follower, and partially covering it. The three-dimensional structure of the wake generated by the follower is similar to those generated by the leaders, with alternating vortex rings that show spanwise compression as highlighted by the black dashed lines. A perspective view of such flow features are also shown for the in-phase (Fig.~\ref{fig:3foils_3Dwake}c) and out-of-phase (Fig.~\ref{fig:3foils_3Dwake}f) cases. Figures~\ref{fig:3foils_2Dwake}a and~\ref{fig:3foils_2Dwake}c show this direct vortex impingement at the midspan ($z=0$). We can observe how the positive (red) vorticity within region A forms at the leading edge of the follower, remains attached to its surface, and connects to the downstream vorticity generated from the wake of leader 1. The importance of the synchronisation between the follower kinematics and the wake is discussed in Section~\ref{sec:FollowerPerf} where the performance of the follower is investigated. Figures \ref{fig:3foils_2Dwake}b and \ref{fig:3foils_2Dwake}d show the time-average vorticity field for the school, and Figures~\ref{fig:3foils_2Dwake}e and~\ref{fig:3foils_2Dwake}f show the time-average streamwise velocity field for the in-phase and out-of-phase cases, respectively. The black and white lines downstream of the follower highlight the mean jets generated by the follower. Region B in Fig.~\ref{fig:3foils_2Dwake}e highlights a region of around 15\% reduction to the time-average streamwise velocity for the in-phase case along the $y=0$ vertical plane. Such a reduced-speed region was predicted by Weihs~\citep{weihs1973hydromechanics} and appears here, indicating that some flowfield features predicted by two-dimensional, potential flow theory are still present for the three-dimensional schooling foils. 

\section{Hydrodynamic performance of the follower}
\label{sec:FollowerPerf}

\begin{figure}
    \centering
    \includegraphics[width=\linewidth]{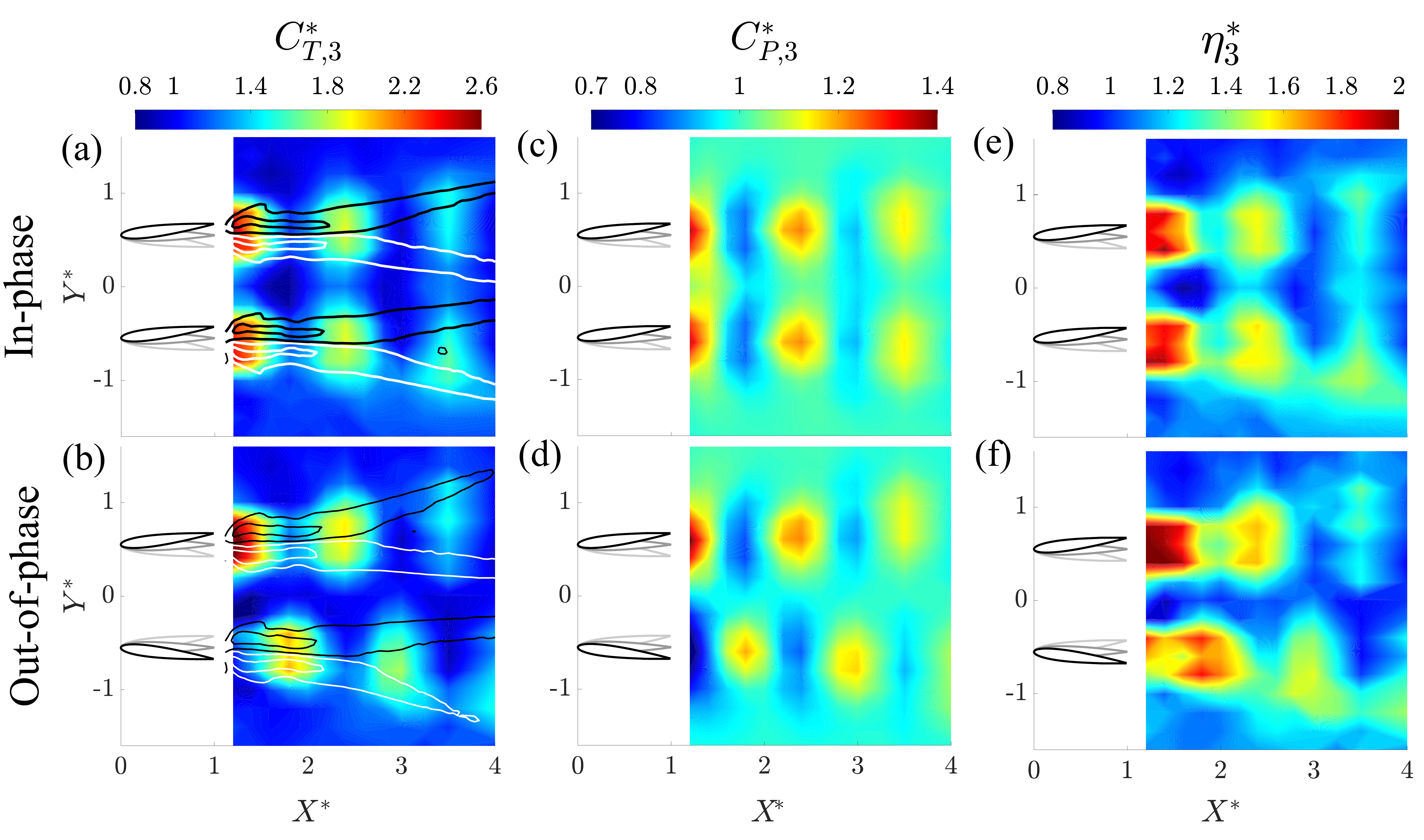}
    \caption{Normalised performance of the follower for the in-phase and out-of-phase cases. (a),(b): Thrust coefficient $C^*_{T,3}$. Black lines are isolines of $\omega_z c/U = 2.5$ and 10; white lines are isolines of $\omega_z c/U = -2.5$ and -10. (c),(d): Power coefficient $C^*_{P,3}$. (e),(f): Efficiency $\eta^*_{3}$.  }
    \label{fig:FollowerPerf}
\end{figure}

In order to better understand the hydrodynamic mechanisms that occur within the school, we turn our attention to the interactions between the leaders' wakes and the follower. Figures \ref{fig:FollowerPerf}a and \ref{fig:FollowerPerf}b present the normalized thrust coefficient of the follower for all spatial arrangements of the school as described in Figure~\ref{fig:domain}. The highest thrust improvement occurs when the follower is within one chord-length downstream of the leaders, with values of $C_{T,3}^* = 2.6$ for the in-phase case and $C_{T,3}^* = 2.9$ for the out-of-phase case occurring at the same location, downstream of leader 1 at $(X^*,\, Y^*) = (1.2,\, 0.6)$. The power coefficient is directly related to the behaviour of the thrust production, with the local minima and maxima of thrust and power nearly coinciding with each other for the two cases studied. Overall, areas of increased thrust also display increased efficiency despite the power expenditure increases (Figures~\ref{fig:FollowerPerf}c and \ref{fig:FollowerPerf}d). Figures~\ref{fig:FollowerPerf}e and \ref{fig:FollowerPerf}f show that the follower achieves the highest measured efficiencies in close proximity to an upstream leader with phase synchrony $\phi=0$. For the in-phase leaders, the peak follower efficiency is $\eta^*_{3,\text{max}}=2$ and occurs downstream of both leader 1 and leader 2. For the out-of-phase leaders the peak efficiency achieved is the same, but only occurs downstream of leader 1 which is in-phase with the follower. For this schooling case (out-of-phase leaders) the second highest peak in efficiency is shifted about $0.4$ chords downstream of leader 2 since the synchrony between leader 2 and the follower is $\phi = \upi$.

\subsection{Reduced flow region and the drafting mechanism}
\label{subsec:3foil_drafting}
With the follower performance presented, we now revisit the mean flow field of the leaders in Figures~\ref{fig:Leaders_2Dwakes}e and~\ref{fig:Leaders_2Dwakes}f, where we see a reduced flow region between the vortical wakes of the leaders, as predicted by \cite{weihs1973hydromechanics}. 
For example, at $(X^*,\, Y^*) = (1.6,\, 0)$ the mean flow speed is 10\% lower than the freestream speed for the in-phase case and 15\% lower for the out-of-phase case. If placed in a steady flow at that reduced speed (in the absence of the time-varying wake structures) the foil is expected to have a 19\% reduction in its viscous drag force for the in-phase case, and a 28\% reduction for the out-of-phase case (assuming a constant drag coefficient and a $U^2$ drag law), which would lead to a higher net thrust relative to our baseline case -- a thrust benefit. However, the follower foil at that location has a 15\% and 4\% \emph{thrust penalty} compared to an isolated foil for the in-phase and out-of-phase cases, respectively. Figure \ref{fig:FollowerPerf} shows that the performance of the follower is not improved along the center line ($Y^* = 0$) in precisely the locations where the follower is postulated to gain a benefit from drafting \citep{weihs1973hydromechanics}. In fact, the data from the three foil school show that the performance benefits are observed in the \textit{accelerated jet flow regions} directly behind the leaders where the drafting hypothesis would predict the greatest performance \textit{penalties} to occur. 

While drafting is a widely accepted mechanism for drag reduction, our results indicate that the vortex interactions outweigh the viscous effects from drafting in this case. Previous studies of in-line formations show that vortex interactions can increase the thrust of a downstream follower even when placed in the leader's accelerated-flow wake that \emph{increases} its time-average viscous drag~\citep{kurt2018flow,Boschitsch2014,ormonde2024two}. Now, our data show that the time-varying effects of the vortex interactions also eclipse the expected thrust benefits due to reduced viscous drag in a reduced-flow wake. The drafting mechanism thus cannot be assumed to improve the performance of oscillating propulsors without explicitly accounting for the significant impacts of the unsteady vortex interactions, as will be shown in the next section. It is worth noting that the observed behaviour may change for leaders placed at different lateral distances or for different pitching amplitudes $A^*$ and Strouhal numbers $St$ since these parameters will directly affect the width of the wakes produced by the leaders such that the interaction of the follower with the wakes at the centre line $Y^*=0$ may change significantly. Additionally, pitching hydrofoils are simple models of only a propulsive fin of a fish, whereas fish have a propulsive fin and a body that mostly generates drag \citep{smits2019undulatory}. With a larger viscous drag source, it might still be possible that fish-like swimmers extract performance benefits from drafting within this reduced flow region. While the present results indicate that the drafting mechanism may not be as important as previously hypothesised, this is still an open question that deserves further investigation.

\subsection{Vortex-body mechanism} \label{subsec:3foil_vortexbody}
The thrust coefficient map (Figure~\ref{fig:FollowerPerf}) shows a very distinctive spatial arrangement with regions of high thrust located within the accelerated jet flow behind the leaders. Specifically, these regions show a subtle double peak structure with the peaks located along the paths of the wake vortices shed by the leaders, represented by the black (white) isolines of positive (negative) time-average vorticity from measurements of just the two leaders alone. The thrust production is thus maximally enhanced by a direct impingement of the oncoming vortices onto the follower as observed in previous work \cite[]{ormonde2024two}. The spatial periodicity of these high-thrust regions coincides with the wake wavelength $\lambda$ of the leader's wake shown in Fig. \ref{fig:Leaders_2Dwakes}a and \ref{fig:Leaders_2Dwakes}b, in agreement with previous findings for 2D and 3D foils located directly downstream of a single in-line fore-foil \citep{kurt2018flow}. The direct vortex impingement changes the instantaneous effective angle of attack of the follower, which in turn changes its cycle-averaged performance. Figure \ref{fig:FollowerPerf} indicates that the same mechanism exists for the wakes shed by the two interacting leaders. Following a direction perpendicular to the vortex trajectories we also see a decay in the thrust of the follower and a decay as the follower is placed further downstream of the leaders.

\begin{figure}
    \centering
    \includegraphics[width=0.8\textwidth]{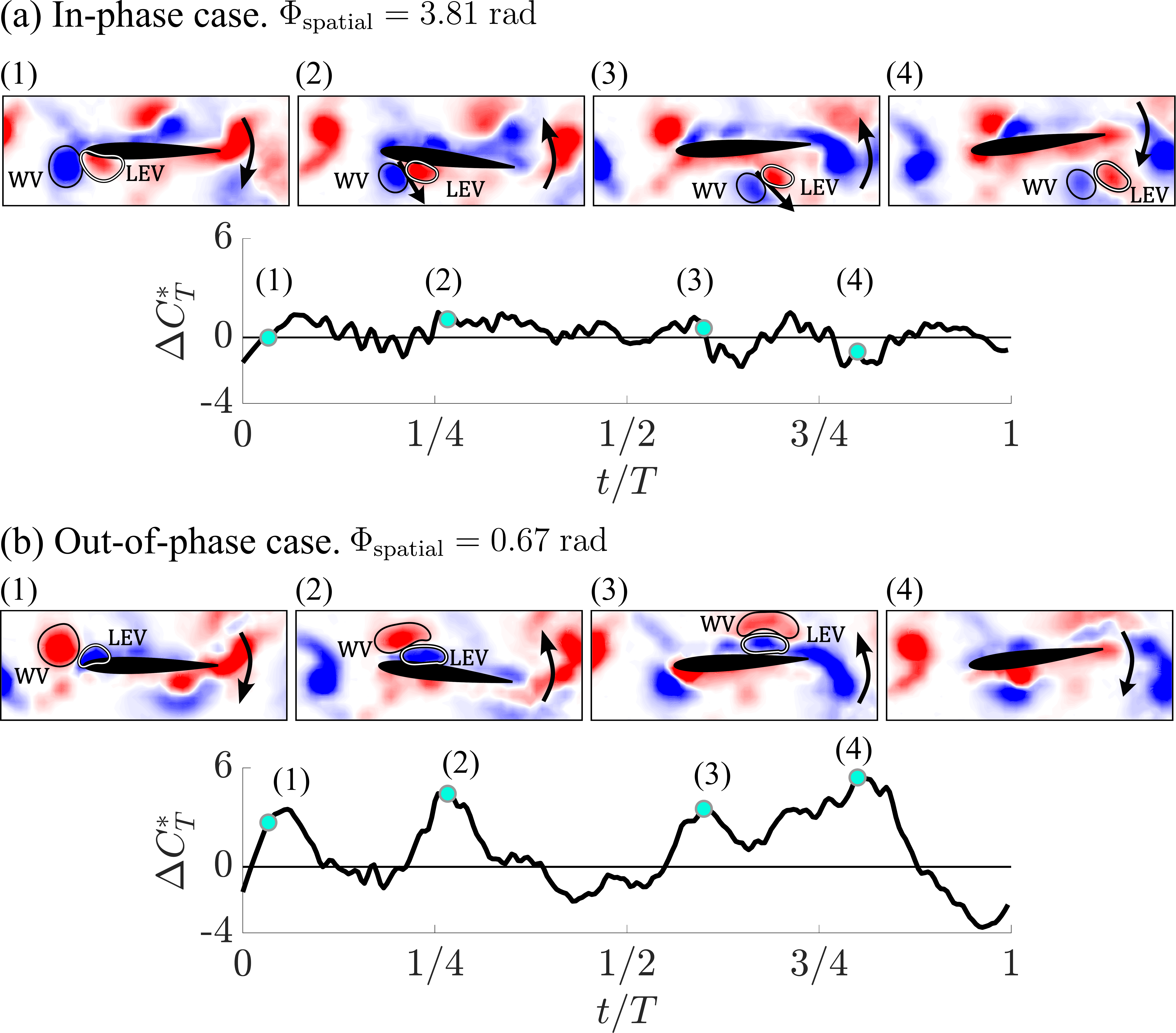}
    \caption{Wake-foil interaction and its impact on the cycle-average thrust production for the follower foil at $(X^*,Y^*) = (1.8,-0.6)$.(a) $\Phi_{\text{spatial}}=3.81$ rad. The interaction of the wake vortex (WV) with the leading edge vortex (LV) causes it to prematurely detach from the foil surface during the cycle. (b) $\Phi_{\text{spatial}}=0.67$ rad. The wake vortex (WV) interacts with the leading edge vortex (LV) such that it remains attached to the foil surface up until it merges with the vorticity shedding from the trailing edge, increasing the thrust generation.
    Non-dimensional time at snapshots (1): $t/T = 1/30$; (2): $t/T = 8/30$; (3): $t/T = 18/30$; (4): $t/T = 24/30$.}
    \label{fig:3foil_vortexImpingement}
\end{figure}

The synchrony between the vortex impingement timing and the foil motion dictates where the peaks are located in the 2D plane. Figure \ref{fig:3foil_vortexImpingement} shows the follower interacting with the wake of leader 2 at the spatial location $(X^*,\, Y^*) = (1.8,\, -0.6)$, for the in-phase case (Figure \ref{fig:3foil_vortexImpingement}a) and the out-of-phase case (Figure \ref{fig:3foil_vortexImpingement}b). The change in the phase-averaged thrust caused by the vortex-foil interaction is plotted as $\Delta C^*_T$, being the difference between the schooling foil and an isolated foil, normalised by the isolated foil as,
\begin{align}
\Delta C^*_T = \left( C_{T,3} - C_{T\text{,iso}} \right) / \ \overline{C}_{T\text{,iso}}. 
\end{align}

Figure~\ref{fig:3foil_vortexImpingement}a shows that for the in-phase case at instant (1) the wake vortex (WV) is at the leading edge, moving towards the lower surface at the beginning of the cycle (downstroke). It then interacts with the leading edge vortex (LEV) at instant (2) and induces a downwards velocity that detaches the LEV from the foil surface during the upstroke at instants (3) and (4). The time-varying thrust is modestly affected by the vortex interaction, and the follower's mean thrust is somewhat elevated from an isolated foil with $C_{T,3} = 1.15$, however, this is lower than the maximum achievable mean thrust at that spatial location. This interaction also leads to a 14\% reduction in power.  

In Figure~\ref{fig:3foil_vortexImpingement}b the follower is at the same spatial location, but is out-of-phase with leader 2, which, consequently, changes the synchrony between the foil motion and the impinging vortex. At instant (1) the wake vortex (WV) impinges onto the follower at the beginning of the cycle (downstroke). The leading edge vortex (LEV) remains attached to the upper surface throughout all of the downstroke, and remains attached during the earlier portion of the upstroke at instant (2) and later in the upstroke at instant (3). At instant (4) the LEV merges with the trailing edge vortex and detaches from the upper surface. Snapshots (3,4) show the next wake vortex (blue) interacting with the next LEV (red) on the lower surface during the upstroke. The time-varying thrust is observed to be significantly enhanced throughout most of the cycle and the mean thrust, $C^*_{T,3} = 2.0$, is twice the thrust of an isolated foil while the mean power also increases by 20\%.

\begin{figure}
    \centering
    \includegraphics[width=0.7\linewidth]{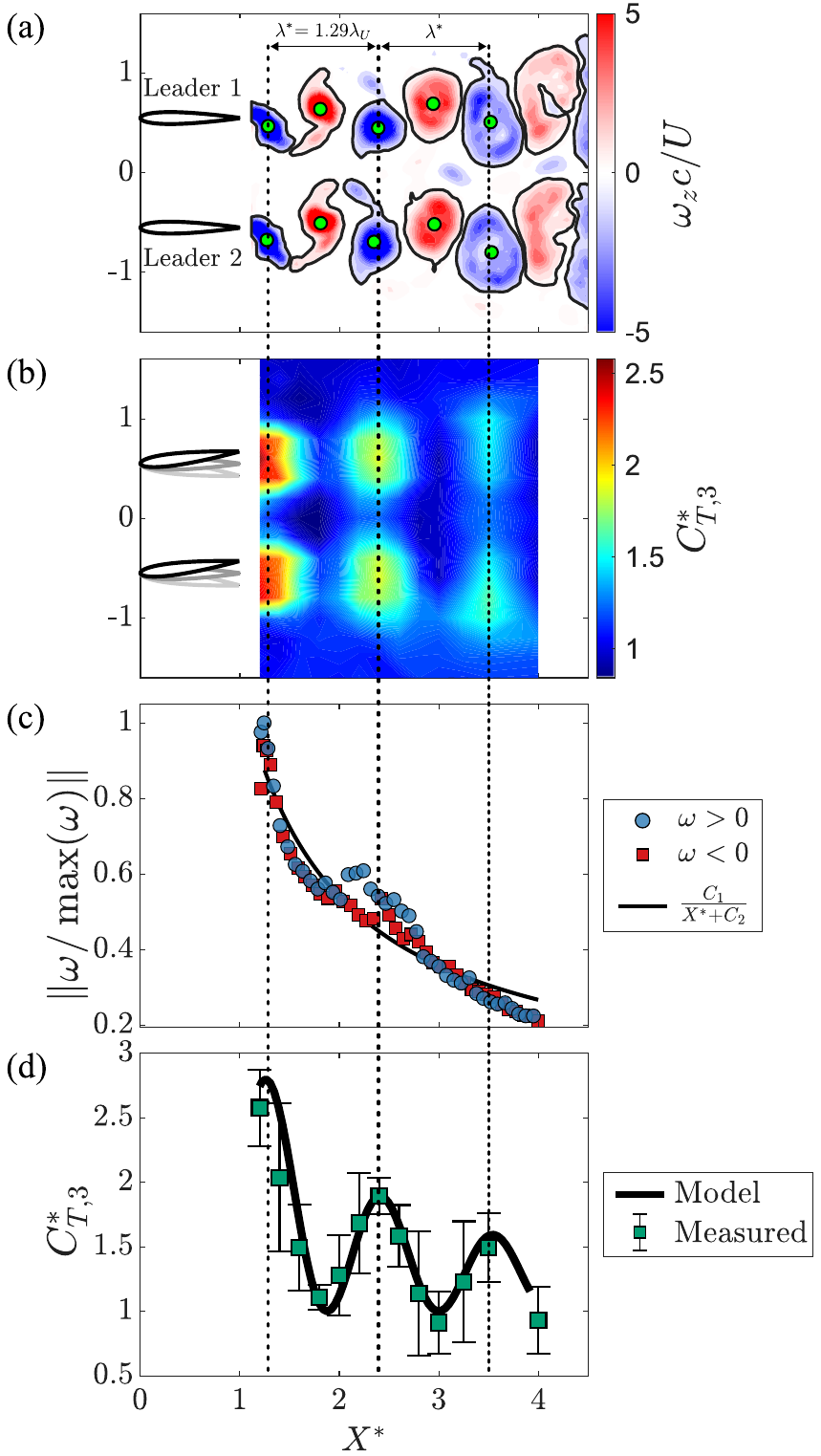}
    \caption{Relationship between the wake dynamics of Leader 1 and the thrust of a follower foil directly downstream of Leader 1, at $(X^*,Y^*)=([1.1 ,\ 4],0.6)$. Results for in-phase leaders.  (a) Measured wake wavelength of the leaders in the absence of a follower, defined as the average distance spanning each set of three consecutive centroids in the streamwise direction. The vortex boundary is defined by the iso-line of constant vorticity (black line) $|\omega|=0.02\max(\omega)$. Snapshot shown for $t/T = 0$. (b) Thrust map of the follower foil highlighting streamwise locations of local maxima, approximately spaced by $\lambda^*$. (c) Streamwise decay of vorticity within each identifiable vortex for all instants of the flapping cycle $t/T=[0,1]$. $\|\omega/\max(\omega) \|= C_1/(X^*+C_2)$, where $C_1=1$ and $C_2=-0.18$. (d) Predictive model of the follower's thrust when directly downstream of the leader $C^*_{T,3} = 1 + C_1/(X^*+C_2)\sin(1+\Phi)$, where $\Phi=2\pi X^*/\lambda^*+\phi$. The vertical dotted lines illustrate how the measured wake wavelength $\lambda^*$ in (a) matches the streamwise spacing of local peaks of thrust in (b) and (d).}
    \label{fig:spatialPhase}
\end{figure}

Descriptions of the synchrony between an impinging wake and a downstream foil have been presented previously \citep{portugal2014upwash, kurt2018flow,Lee2023} and described in terms of a spatial phase defined as $\Phi_U=2\upi f\,X/U + \phi$, taking the freestream (or swimming) speed as the advection velocity of the wake vortices. We propose an improved definition for the spatial phase utilising the \emph{measured} wake wavelength $\lambda$ instead of the estimated value of $\lambda_U = U/f$ based on the freestream speed. An average wake wavelength is measured for the two leaders in the absence of a follower. The measured wavelength is defined as the mean distance along the streamwise direction between the first and third centroids of vorticity within successive triplets of shed vortices, as shown in figure~\ref{fig:spatialPhase}(a).  

The proposed spatial phase $\Phi$ is thus defined as
\begin{equation}
    \Phi = \frac{2\pi}{\lambda^*} X^* + \phi 
\end{equation}
\noindent where $\lambda^*=\lambda/c$ is the dimensionless \textit{measured} wake wavelength, and $\phi$ is the temporal synchrony between the leader whose wake is impinging on the follower. Our data shows that the measured wavelength is 29\% larger than the estimated $\lambda_U$ based on the free-stream speed. This discrepancy is expected due to the fact that the pitching foils produce a net thrust force, thus creating a jet of downstream accelerated flow as shown in figure~\ref{fig:Leaders_2Dwakes}(e) and (f). This region of accelerated flow directly impacts the advection speed of the wake vortices, stretching it along the streamwise direction compared to the free-stream based wavelength prediction.

Our goal now is to connect the properties of the wake shed by the leaders, without the follower present, to the performance of the follower foil when it is located directly downstream of them. Figure~\ref{fig:spatialPhase}b shows the streamwise distance between thrust peaks of the follower, which are approximately separated by a distance of $\lambda^*$. A simple model of the follower's thrust is postulated to be proportional to a periodic function of the measured spatial phase $\Phi$ such as $C^*_{T,3} \propto \sin{\left(\Phi\right)}$ and to have a streamwise decay directly tied to the decay of vorticity in the wake. The streamwise decay of vorticity in the wake of the leaders (in the absence of a downstream follower) is due to the spreading of vorticity as the vortices advect downstream and is shown in Figure~\ref{fig:spatialPhase}c. The average vorticity contained within each vortex is measured for all instants $t/T=[0,1]$ as a function of the vortex centroid position. The measured magnitude of positive (red) and negative (blue) vortices decay downstream, and is well described by an inverse function of the form $\|\omega/\max(\omega)\|=C_1/(X^*+C_2)$, where $C_1=1$ and $C_2=-0.18$. The hyperbolic decay used in our model is inspired by the decay law predicted by Lamb's vortex model as described by \cite{Ponta2010vortex}, which was observed in their measurements of a von-Karman cylinder wake. Our proposed model for the thrust of the follower is given by 
\begin{equation}
    \Delta C^*_{T,3} = 1+\frac{C_1}{X^*+C_2} \left(1+ \sin{\Phi}\right), 
    \label{eqn:Phi}
\end{equation}
\noindent with $C_1 = 1$, $C_2=-0.18$ and $\Phi = 2\pi f/1.29U X^* + \phi$ for the follower directly behind one of the leaders. The term $2\upi f/1.29 U$ accounts for the measured wake wavelength downstream of Leader 1. The model successfully predicts the streamwise behaviour of the thrust coefficient of the follower as shown in Figure~\ref{fig:spatialPhase}d. The spatial oscillations are primarily due to the spatial phase $\Phi$, which governs how the wake vortex (WK) interacts with the follower's leading edge vortex (LEV) as depicted in Figure~\ref{fig:3foil_vortexImpingement}, while the streamwise wake strength decay affects the effective angle of attack of the follower foil and is a function of the distance $X^*$ from the follower to the trailing edge of the leader. 

The spatial phase has been previously defined throughout literature using the estimated (approximate) wake wavelength involving the freestream speed $U$ and the downstream distance $x$ \citep{li2020vortex,portugal2014upwash,newbolt2019flow,heydari2021school}. This is a suitable approximation as long as the wake follows a straight path in the $x$-direction and its advection speed is close to $U$ in magnitude and remains constant as it advects downstream. For the wakes presented here, however, these assumptions do not hold true. The spatial phase parameter has also been used to study arrays of tandem oscillating foils for energy harvesting. A similar correction of the wake advection speed has been shown to improve the prediction of the optimal spacing for the performance of the downstream foil~\citep{Handy-Cardenas2025optimal,ribeiro2021wake} compared with predictions based on an uncorrected wake advection velocity equal to the freestream speed~\citep{KinseyDumas2012optimal}. Unlike the wakes discussed in this study, which produce regions of mean accelerated flow (see Figures~\ref{fig:Leaders_2Dwakes}c and~\ref{fig:Leaders_2Dwakes}f) and a reverse von K\'arm\'an vortex street, the wakes generated by drag-producing energy-harvesting foils exhibit a mean flow deficit and different vortical structures. Despite this difference, the spatial phase parameter encapsulates the combined effects of temporal synchronization and spatial separation between two tandem foils into a single parameter. The spatial phase describes the timing between the downstream foil kinematics and the impingement of the wake shed by the upstream foil.

The model's insight is that \textit{only the wake information from the leaders alone} is needed to accurately predict the optimal vortex-body phase synchronization of downstream followers. Therefore a school with optimal phase synchronization of any size can be designed by progressively marching from leader to follower in the following way. First, the wake wavelength for a leader's wake can be calculated/measured and then the optimal phase of a follower placed in this flow can be chosen. Then the wake wavelength from the leader-follower pair can be calculated/measured and another follower can be placed downstream with an optimal synchronization, and so forth. In this way a school of $N$ swimmers with optimal synchronization can be designed by progressively placing  followers downstream and choosing their optimal synchronization based on $N$ simulations or experiments of the leader's wake flow rather than $N\times M$ simulations or experiments when $M$ is some number of temporal phases ranging from $0$ to $2\pi$.

\section{Hydrodynamic performance of the leaders}
\label{sec:LeaderPerf}
\begin{figure}
    \centering
    \includegraphics[width=\linewidth]
    {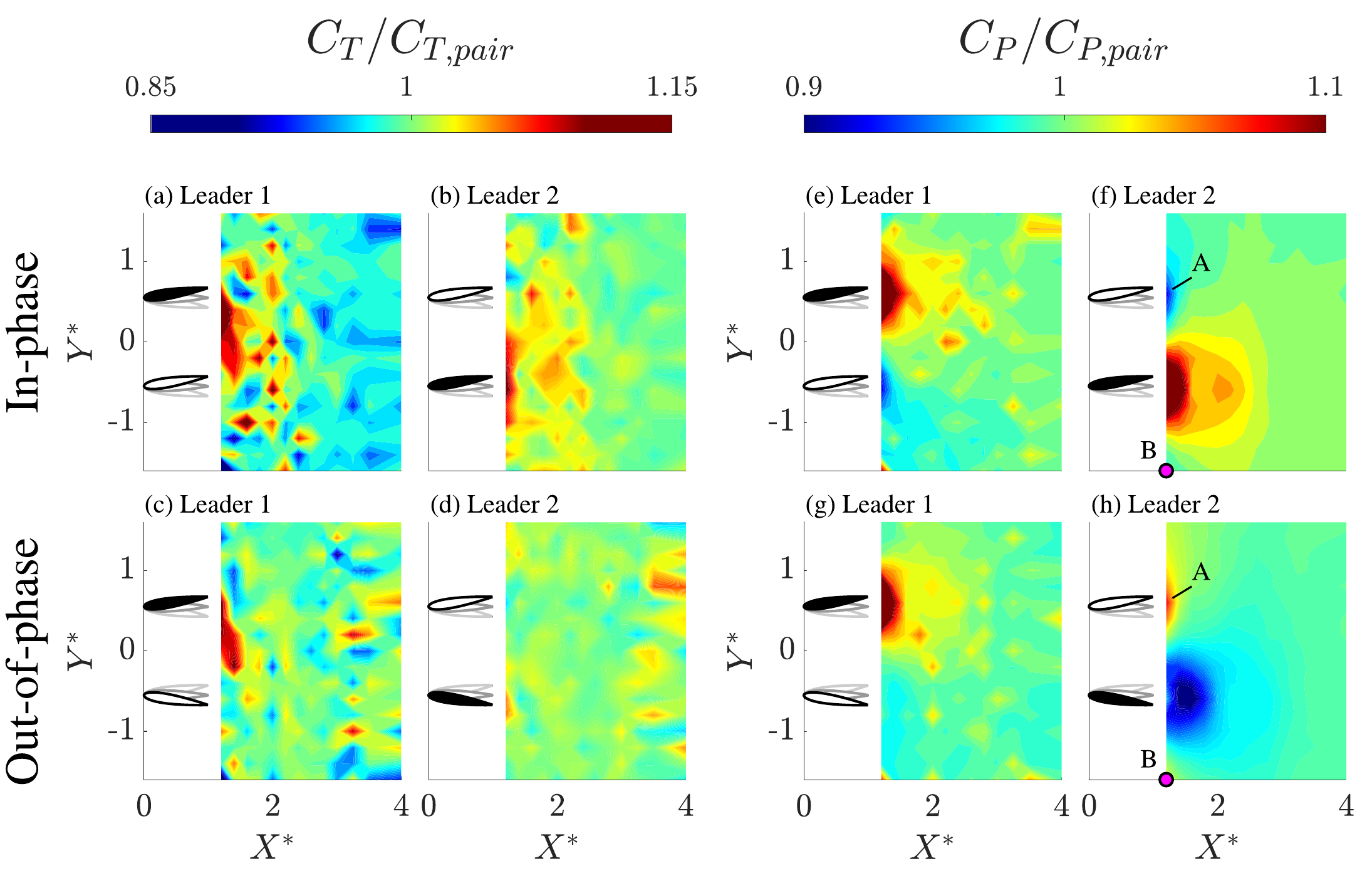}
    \caption{Performance landscape of leader 1 and leader 2 as a function of the position of the follower. The black-coloured foil indicates the foil to which the figure corresponds to. Thrust and power coefficients are normalized by the performance of each respective leader in the side-by-side arrangement in the absence of a follower. (a),(b): Thrust of leader 1 and leader 2, respectively, for an in-phase synchrony. (c),(d): Thrust of leader 1 and leader 2, respectively, for an out-of-phase synchrony. (e),(f): Power of leader 1 and leader 2, respectively, for an in-phase synchrony. In (f), points A and B are equidistant from leader 2. (g),(h): Power of leader 1 and leader 2, respectively, for an in-phase synchrony. In (h), points A and B are equidistant from leader 2.}
   \label{fig:3foil_PerfLeaders}
\end{figure}

The performance of the leaders is affected by the follower when it is within approximately one chord-length away from either leader 1 or leader 2. Figure \ref{fig:3foil_PerfLeaders} shows the thrust and power maps for leader 1 and leader 2 as a function of the follower's position. The effect of a downstream follower on a single upstream foil has been extensively documented in the literature \citep{Boschitsch2014,kurt2018flow} and is also observed in our results for two side-by-side leaders. Figure~\ref{fig:3foil_PerfLeaders}a shows, for the in-phase case, that leader 1 produces higher thrust when the follower is directly interacting with its wake. The same is true for leader 2 for the in-phase case shown in Figure~\ref{fig:3foil_PerfLeaders}b, denoted by the deep red region directly behind leader 2 in the figure. For the out-of-phase case, leader 1 produces higher thrust when the follower is in close proximity (Figure~\ref{fig:3foil_PerfLeaders}c). However, for the out-of-phase case, Figure~\ref{fig:3foil_PerfLeaders}d shows no changes in the thrust of leader 2 when the follower is in its close vicinity. 

The power coefficient of the leaders as a function of the position of the follower is presented in Figures~\ref{fig:3foil_PerfLeaders}e\textendash f. For the in-phase case, Figure~\ref{fig:3foil_PerfLeaders}e shows that leader 1 experiences a 10\% \emph{increase} in its power expenditure when the follower is directly behind it (deep red region at $1.2\leq X^*\leq 1.8$ and $Y^*=D^*/2$). Conversely, when the follower is behind of leader 2, leader 1 experiences a 5-10\% \emph{decrease} in its power expenditure as shown by the deep blue region at $1.2\leq X^* \leq 1.4$ and $Y^*=-D^*/2$. Figure~\ref{fig:3foil_PerfLeaders}f shows a symmetric pattern in the power coefficient map of leader 2: The power expenditure of leader 2 increases when the follower is closely behind it (deep red region at $1.2 \leq X^* \leq 1.8$ and $Y^*=-D^*/2$), and decreases when the follower is closely behind of leader 1 (deep blue region at $1.2\leq X^* \leq 1.4$ and $Y^*= D^*/2$). 

The performance maps for the thrust and power of the two leaders shown in Figure~\ref{fig:3foil_PerfLeaders} reveal how performance changes can propagate upstream along the school for sufficiently compact formations. They also show that the temporal synchronization between foils is an important parameter dictating the impact that the downstream foil has on its upstream neighbours, something that has been previously established for in-line formations~\citep{kurt2018flow,ramananarivo2016flow} and has now been experimentally confirmed for a larger sized school of three foils and several different spatial formations. What has \emph{not} been observed in previous work are indirect, or ``cascading'' effects where one foil affects their nearest upstream neighbour (second foil), which in turn affects a third foil. Figures~\ref{fig:3foil_PerfLeaders}f and \ref{fig:3foil_PerfLeaders}h show this phenomenon for the power coefficient of leader 2. Points A and B are equidistant from the trailing edge of leader 2. When the follower is positioned at point B (pink circle), below leader 2, there is no significant change in the power coefficient of leader 2 (black foil). However, when the follower is placed at point A, above leader 2, the data shows a 5\% decrease in the power coefficient of leader 2 for the in-phase case (Figure~\ref{fig:3foil_PerfLeaders}f) and a 10\% increase for the out-of phase case (Figure~\ref{fig:3foil_PerfLeaders}h). We hypothesize that this is caused by both a weak direct effect from the follower on leader 2 and an indirect effect where the proximity of the follower to leader 1 alters the bound circulation of leader 1, which in turn affects its interaction with leader 2. So, in essence the upstream effect of the follower onto leader 1 is propagated to leader 2. Further studies investigating this phenomenon, especially with more foils interacting, are needed to clarify this hypothesis.

\section{Collective performance, fluid mediated forces, and formation stability}
\label{sec:3foilSchool}
We now shift our focus to the collective performance and stability of the school. On the previous sections, we established how each foil is affected by the school's spatial formation and the temporal synchronization of the three foils. Here, we present maps of the collective performance of the school and also analyse the stability of those formations. We then discuss the relationship between high-performance formations and school stability.

\subsection{Collective performance}
\label{subsec:CollectivePerf}

\begin{figure}
    \centering
    \includegraphics[width=\linewidth]{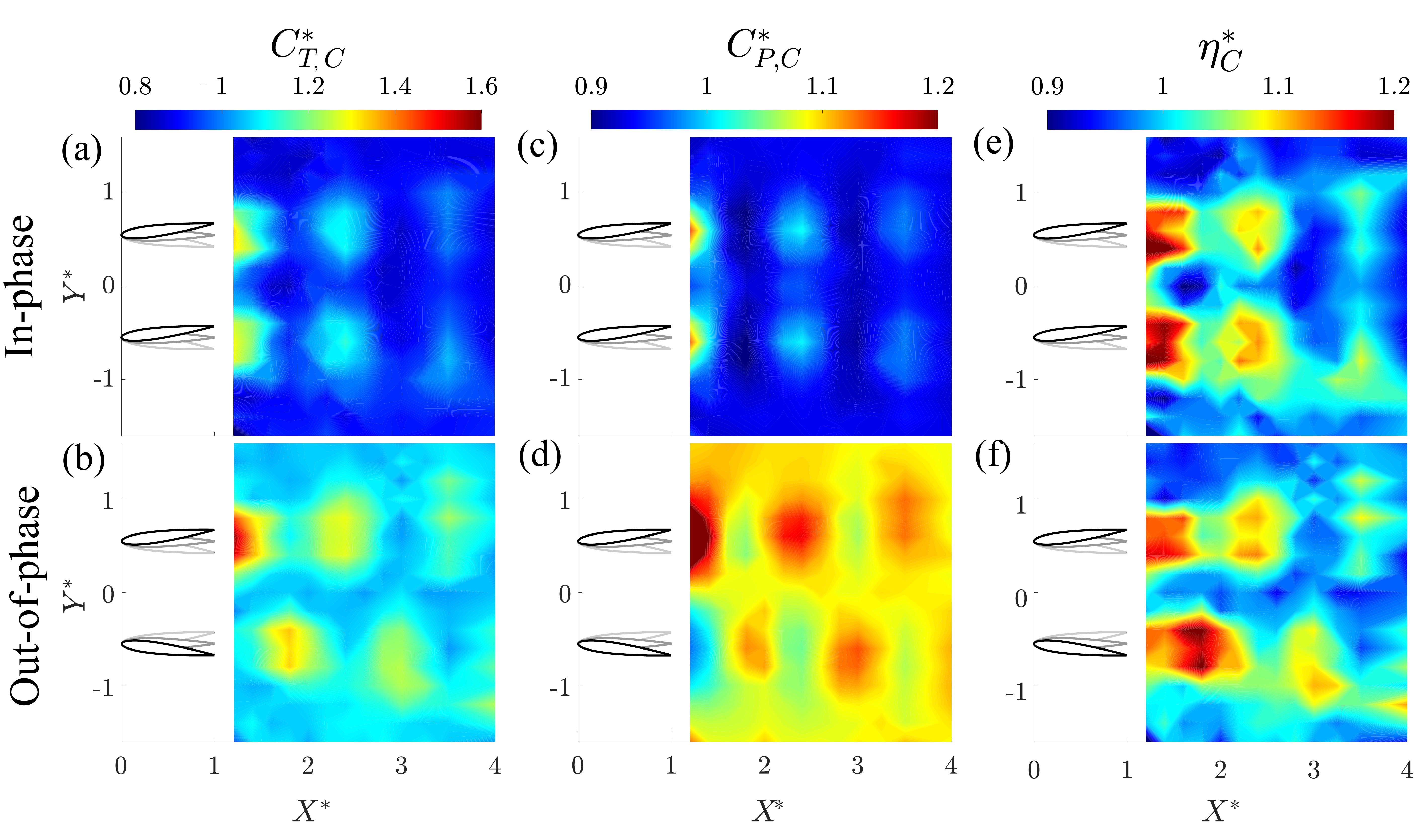}
    \caption{Normalised collective performance for the in-phase and out-of-phase cases. (a),(b): Thrust coefficient $C^*_{T,C}$. (c),(d): Power coefficient $C^*_{P,C}$. (e),(f): Efficiency $\eta^*_{C}$.  }
    \label{fig:CollecPerf}
\end{figure}

The collective thrust, power, and efficiency of the school are presented in Figure \ref{fig:CollecPerf} for the in-phase and out-of-phase cases. High collective performance improvements are achieved for both cases, with the peak collective normalised thrust of $C^*_{T,C} = 1.34$ at $(X^*,\,Y^*) = (1.2,\,0.6)$ for the in-phase case and $C^*_{T,C} = 1.58$ at $(X^*,\,Y^*) = (1.2,\,0.6)$ for the out-of-phase case. The out-of-phase case shows thrust enhancement for almost all the formations investigated since the two leaders by themselves already experience a thrust improvement. For the in-phase case this is not observed, and the minimum thrust value achieved is $C^*_{T,C} = 0.8$. This number is largely due to the baseline thrust produced by the two in-phase leaders. In the absence of a follower the thrust coefficient of two in-phase leaders is 17\% lower than the baseline case for an isolated swimmer.

The streamwise spatial distribution of power and efficiency peaks coincide with the streamwise locations of peak thrust. While the power requirements increase for higher-thrust formations, Figures~\ref{fig:CollecPerf}e\textendash f show that the normalised efficiency is still greater than one, indicating an efficiency enhancement over an isolated foil. Overall, high-thrust formations yield higher propulsive efficiencies. Peak efficiencies of $\eta^*_C = 1.24$ are achieved at $(X^*,\,Y^*) = (1.4,\,0.4)$ for the in-phase case (Figure~\ref{fig:CollecPerf}e)  and $\eta^*_C = 1.21$ at $(X^*,\,Y^*) = (1.8,\,-0.8)$ for the out-of-phase case (Figure~\ref{fig:CollecPerf}f). The highest propulsive performances are found for compact arrangements, where the follower is positioned within one chord-length of the trailing edge of either of the leaders.

\subsection{Lateral forces and stability considerations}
\label{subsec:stability}
\begin{figure}
    \centering
    \includegraphics[width=\linewidth]{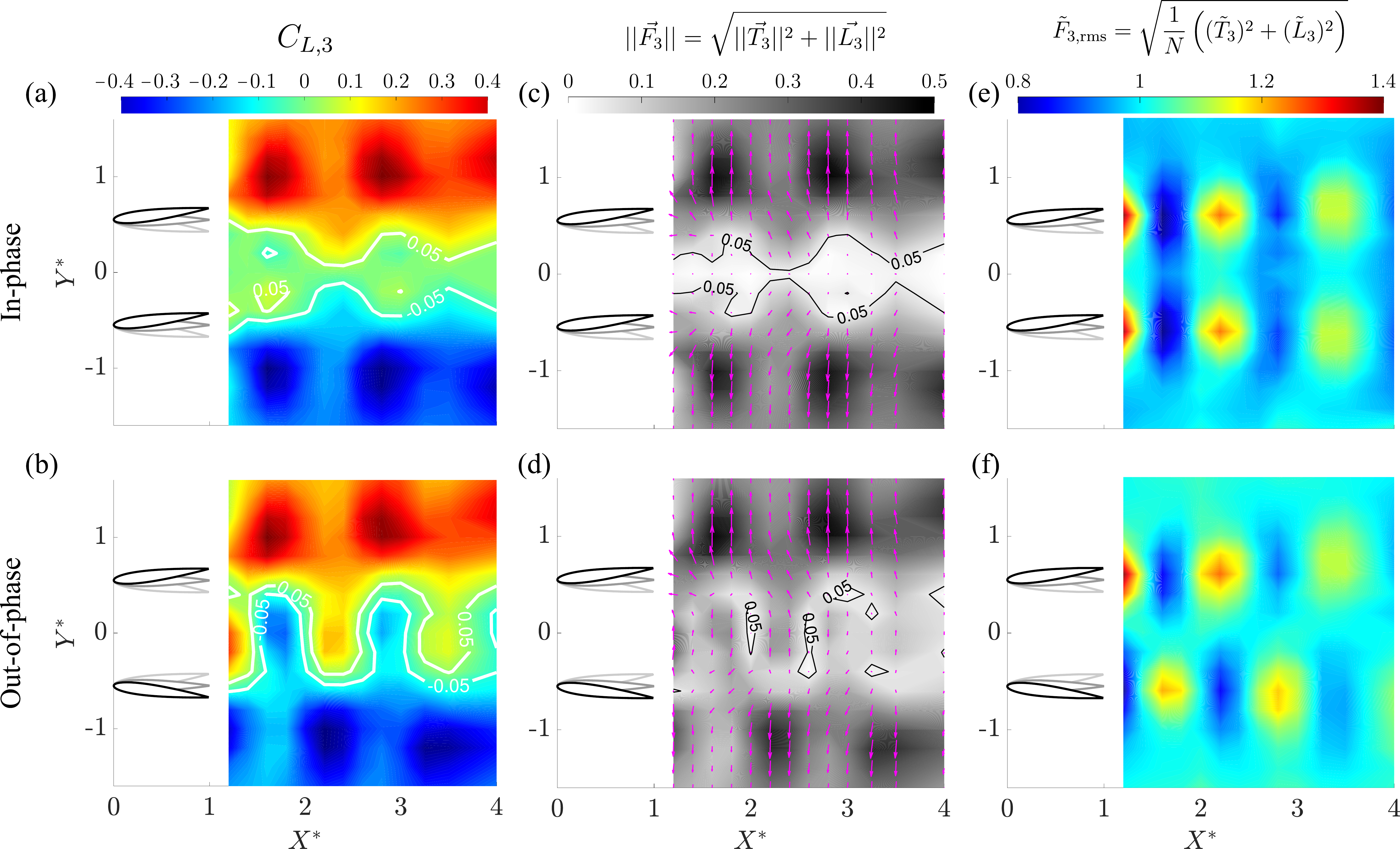}
    \caption{(a) and (b): Time-average lift coefficient of the follower $C_{L,3}$ for the in-phase and out-of-phase cases, respectively. The white lines are iso-lines of $\left|C_{L,3} \right|  = 0.05$. (c) and (d): Time-average vector fields of a schooling follower compared to the isolated case $\vec{F}_{\Delta}(x,y)$ for the in-phase and out-of-phase cases, respectively. Vectors represent the direction and magnitude of the time-average force produced by the follower at each location $\vec{F_3} = T\hat{x} + L\hat{y}$. The black solid lines are iso-lines of $F_{\Delta} = 0.05$. (e) and (f): Root mean square fluctuations of the resultant vector $\mid \Vec{F_{\Delta}}(x,y) \mid$ for the in-phase and out-of-phase cases, respectively.}
    \label{fig:CL_follower}
\end{figure}

Figures~\ref{fig:CL_follower}a and~\ref{fig:CL_follower}b show the time-average lift coefficient of the follower as a function of its position for the in-phase and out-of-phase cases, respectively. For the in-phase case, the follower produces small lift forces of less than $\left|C_L \right| < 0.05$ within a relatively narrow region, highlighted by the white isolines in Figure~\ref{fig:CL_follower}a. Higher lift forces are produced outside of this region. At $Y^* > D^*/2$, regions of positive (upwards) lift up to $C_L \approx 0.45$ are observed, and regions of negative (downwards) lift of the same magnitude can be found at $Y^* < -D^*/2$. Based on the follower's lift coefficient map, an unconstrained oscillating foil located anywhere in the domain at $Y^*>D/2$ or $Y^*<-D/2$ will experience a net lateral force that pushes it \emph{away} from the two upstream leaders. As a result, the follower is required to alter its kinematics in order to maintain its lateral position relative to the leaders, counterbalancing the flowfield effects caused by its upstream neighbours. On the other hand, if the follower stays within the region bounded by the white iso-lines, it will experience very small lift forces. As a consequence, significantly less control input is required from a follower to maintain its lateral position within that region. Figure~\ref{fig:CL_follower}b shows the same map for the out-of-phase case. Similarly to the in-phase case, high-magnitude lift forces of up to $\left|C_L \right| \approx 0.45$ are produced by the follower in the two regions of $\left|Y^*\right|>D^*/2$. However, in-between the two leaders at $\left|Y^*\right|<D^*/2$, the lift forces only stay below the 0.05 threshold within a very small, sinuous region that varies quickly along the streamwise direction, as highlighted by the white isolines. As a result, Figures~\ref{fig:CL_follower}a and~\ref{fig:CL_follower}b suggest that, compared to the in-phase leaders that produce a region of low-magnitude lift forces, the two out-of-phase leaders require significantly more input control/corrections from the follower to maintain a given spatial formation downstream of its upstream neighbours.

Next, in Figures~\ref{fig:CL_follower}c and~\ref{fig:CL_follower}d we combine the lift and thrust forces of the schooling follower into a resultant vector and compare it to a follower in isolation. The resulting vector field describes the the change in the resultant force produced by the schooling follower relative to the isolated case. The resultant force vector $\vec{F_{\Delta}}(x,y)$ is defined as 
\begin{equation}
    \frac{ \vec{F}_{\Delta}}{1/2 \rho U^2 c s }(x,y) = \left( C_{T,3}(x,y) - C_{T\text{,iso}} \right)\hat{x} + \left( C_{L,3}(x,y) - C_{L\text{,iso}} \right)\hat{y}
\end{equation}
\noindent where $C_{T\text{,iso}}$ is the isolated thrust coefficient and $C_{L\text{,iso}}$ is in principle zero, given its symmetric motion. A small lift bias $C_{L,3}^{\text{iso}} = -0.04$ was measured for the isolated follower, however, due to small foil misalignments and other unknown experimental errors. 

Figures~\ref{fig:CL_follower}c and~\ref{fig:CL_follower}d show the force vector fields for the in-phase and out-of-phase cases, respectively, where the background contour is the magnitude of the force vector $\left|\vec{F}_{\Delta}(x,y)\right|$, used to help visualize the magnitude of the magenta-coloured vectors. Isolines of value $F_{\Delta} = 0.05$ are also used to indicate regions of low change in the total force. It becomes apparent, from the vector fields, that the schooling follower experiences substantially larger changes in its force production in the lateral direction (lift) compared to the streamwise direction (thrust/drag). Figures~\ref{fig:CL_follower}a-d suggest that, in order to stay close to the two side-by-side leaders at the regions $\left| Y^* \right| > D^*/2$, a follower is required to alter its kinematics compared to an isolated foil. Otherwise, the downstream foil will be laterally pushed away from the leaders until it is effectively no longer schooling, i.e., it is no longer hydrodynamically interacting with its neighbours. In order to produce the required time-averaged zero lift required to maintain group cohesion, a self-propelling follower could possibly alter its mean angle of attack (or yaw angle), adopt an asymmetric oscillatory motion, or in the case of a fish or a fish-like robot, employ their control surfaces such as pectoral/dorsal fins to counteract the lift force produced by the caudal fin. In addition, Figures~\ref{fig:CL_follower}c and~\ref{fig:CL_follower}d clearly highlight that formations of high collective performance (see Figure~\ref{fig:CollecPerf}) are not \emph{stable}, since the higher thrust production of the follower is associated with large lift forces. As a consequence, in order for a freely swimming follower to extract large performance benefits from the wake of its upstream neighbours, it is required to counter-balance these lift forces at the expense of some hydrodynamic performance, which has not been accounted for in this study.

The root mean square (rms) of the resultant force fluctuations are presented in Figures~\ref{fig:CL_follower}e and~\ref{fig:CL_follower}f. The thrust and lift fluctuations are calculated by subtracting the mean values from their corresponding time-varying signals $\Tilde{T} = T(t) - \overline{T}$ and lift $\Tilde{L} = L(t) - \overline{L}$. The resultant force is the vector $\mathbf{\Tilde{F}} = \Tilde{T}\hat{x} + \Tilde{L}\hat{y}$, so the rms of the resultant fluctuations is calculated as $\Tilde{F}_{\text{rms}} = \sqrt{\frac{1}{N} \left( \Tilde{T}^2 + \Tilde{L}^2 \right) }$, where $N$ is the number of samples of the time series. The rms values are then normalized by the rms of an isolated foil, measured to be  $F_{\text{iso}}^{\text{rms}} = 0.43$ N. There is a direct correlation between the fluctuations of Figures~\ref{fig:CL_follower}e and~\ref{fig:CL_follower}f, and the follower's coefficients of thrust and power as shown in Figures~\ref{fig:FollowerPerf}a-d. The similarities with the power coefficient are particularly evident, since regions of high and low $C_P$ coincide almost exactly with the spatial distribution of the fluctuations for both synchrony cases. The same overlap cannot be said about the spatial distribution of the lateral forces, since the peaks of lift forces from Figures~\ref{fig:CL_follower}a and~\ref{fig:CL_follower}d do not overlap with the peaks of force fluctuations. The data indicate that the force fluctuations of the follower directly impact its power expenditure, but are not directly correlated to large mean lateral forces. High mean thrust formations, on the other hand, seem to correlate with increased force fluctuations.  

Consider an unconstrained follower swimming downstream of the two leaders. In order for it to gain energetic benefits from the wake of the leaders, the constrained performance data presented suggests it will have to exert some effort to remain at a high-performance position. For example, in the out-of-phase case the follower must counteract a mean lateral force that will push it away from the $Y^* = 0$ line in between the two leaders. The large force fluctuations detected in the high-performance regions also mean that the follower would likely experience large amplitude recoil motions.  Some regions provide improved thrust and efficiency with increased fluctuations, while others are more stable and should require little control of a follower to maintain its position. The force maps from Figure~\ref{fig:CL_follower}, together with the performance maps from Figure~\ref{fig:FollowerPerf}, thus provide a framework to navigate through different spatial formations downstream of the leaders that generally offer a trade-off between efficiency and stability.

\section{Conclusions}
\label{sec:conclusions}
The present study examines the performance and flow structures of three, three-dimensional pitching hydrofoils across a range of formations that examine the ``back-half" of the classical diamond formation with both in-phase and out-of-phase synchronization between the leaders. The mechanisms of drafting, vortex-body interactions, body-to-body interactions, and wake breakdown were probed for their role in driving the performance of the collective.

The drafting hypothesis proposed by Weihs \citep{weihs1973hydromechanics,weihs1975some} was shown to play no role in the performance benefits of the three foil school.  Reduced flow regions of up to a 17\% reduction in the free-stream speed were identified between the rows of vortices shed from the leaders, as predicted by Weihs. However, the reduced flow regions showed no accompanying regions of increased follower net thrust (reduced drag) when the follower was placed there. In fact, the highest thrust benefits of the follower (190\% more thrust than an isolated foil) occurred directly in the \textit{accelerated} jet flow regions within the rows of vortices shed by the leaders, where the drafting theory would predict a drag increase and thus a thrust penalty.

The follower foil experienced the largest performance variations of all three foils, achieving high performance for formations where the wake of a leader directly impinges onto it. A new spatial phase definition was proposed to identify the optimal synchrony between the foil oscillation and the oncoming wake, as well as the downstream decay in thrust with increasing distance between the follower and the upstream leader. The key features of the model are the use of the actual wake wavelength instead of the estimated wavelength previously used in literature, as well as the wake's streamwise vorticity decay. The optimal spatial phase that maximizes the efficiency represents the synchronization between the foil's motion and an impinging vortex such that the oncoming vortex acts to maximize the angle of attack of the follower. A leading-edge vortex is also produced by the follower, which pairs with the impinging vortex to propagate along the foil and merge with the forming trailing-edge vortex.

The follower foil also affects the upstream leader foils when it's sufficiently close to them. It is observed that not only can the follower affect its nearest neighbour, but it triggers a ``cascading'' effect where the nearest leader affects the other leader as well. To the authors knowledge, this is the first time such a phenomenon is reported for experiments with oscillatory foils. Additionally, the thrust data for the leaders suggests that the side-by-side formation is more sensitive to a downstream follower than a single upstream swimmer. We hypothesise that this occurs due to the fact that when the follower disrupts the wake from one leader, the wake from the second leader is also affected, slightly altering the thrust production of both leaders.

The collective performance of the school shows high maximum thrust and efficiency gains for formations where the follower is in close proximity to the leaders and when the follower is in the wake of either of the two leaders. Collective thrust and efficiency enhancements of up to 58\% and 24\% higher than an isolated foil, respectively, are observed.

The lateral forces of the follower were also examined. The magnitude of the changes in the lift forces are significantly larger than the changes in thrust, and the regions of highest thrust coincide with high net lift forces. A narrow region of reduced total lift and thrust changes were found for both in-phase and out-of-phase leaders. This suggests that schooling fish that want to remain in a certain formation may be faced with a dilemma: regions of high performance may require strong lateral control measures and constant trajectory corrections, whereas regions of relative ``calm", or low lateral forces, are not as energetically beneficial. 

In summary, it was discovered that (1) drafting provided no performance benefits despite the presence of reduced flow regions, (2) wake breakdown did not eliminate performance benefits from vortex-body mechanisms within three chord lengths downstream of the leaders, (3) vortex-body interactions drove the performance benefits of the follower and collective, (4) body-to-body interactions played a role for compact formations with the follower less than one chord length downstream of the leader, and (5) the highest collective performance benefits occurred for compact formations.

\bibliographystyle{jfm}
\bibliography{references}

@article{KinseyDumas2012optimal,
    author = {Kinsey, Thomas and Dumas, Guy},
    title = {Optimal Tandem Configuration for Oscillating-Foils Hydrokinetic Turbine},
    journal = {Journal of Fluids Engineering},
    volume = {134},
    number = {3},
    pages = {031103},
    year = {2012},
    month = {03},
    issn = {0098-2202},
    doi = {10.1115/1.4005423},
    url = {https://doi.org/10.1115/1.4005423}
}

@article{Handy-Cardenas2025optimal, 
title={Optimal kinematics for energy harvesting using favourable wake–foil interactions in tandem oscillating hydrofoils}, 
volume={1012}, 
DOI={10.1017/jfm.2025.10198}, 
journal={Journal of Fluid Mechanics}, 
author={Handy-Cardenas, Eric E. and Zhu, Yuanhang and Breuer, Kenneth S.}, 
year={2025}, 
pages={A23}
}

@article{Ponta2010vortex,
    author = {Ponta, Fernando L.},
    title = {Vortex decay in the Kármán eddy street},
    journal = {Physics of Fluids},
    volume = {22},
    number = {9},
    pages = {093601},
    year = {2010},
    month = {09},
    issn = {1070-6631},
    doi = {10.1063/1.3481383},
    url = {https://doi.org/10.1063/1.3481383}
}

@article{newbolt2024flow,
  title={Flow interactions lead to self-organized flight formations disrupted by self-amplifying waves},
  author={Newbolt, Joel W and Lewis, Nickolas and Bleu, Mathilde and Wu, Jiajie and Mavroyiakoumou, Christiana and Ramananarivo, Sophie and Ristroph, Leif},
  journal={Nature communications},
  volume={15},
  number={1},
  pages={3462},
  year={2024},
  publisher={Nature Publishing Group UK London}
}

@article{han2022propulsive,
  title={Propulsive performance and vortex wakes of multiple tandem foils pitching in-line},
  author={Han, Pan and Pan, Yu and Liu, Geng and Dong, Haibo},
  journal={Journal of Fluids and Structures},
  volume={108},
  pages={103422},
  year={2022},
  publisher={Elsevier}
}

@article{streitlien1996efficient,
  title={Efficient foil propulsion through vortex control},
  author={Streitlien, Knut and Triantafyllou, George S and Triantafyllou, Michael S},
  journal={Aiaa journal},
  volume={34},
  number={11},
  pages={2315--2319},
  year={1996}
}

@article{beal2006passive,
  title={Passive propulsion in vortex wakes},
  author={Beal, David N and Hover, Franz S and Triantafyllou, Michael S and Liao, James C and Lauder, George V},
  journal={Journal of fluid mechanics},
  volume={549},
  pages={385--402},
  year={2006},
  publisher={Cambridge University Press}
}

@article{liao2004neuromuscular,
  title={Neuromuscular control of trout swimming in a vortex street: implications for energy economy during the Karman gait},
  author={Liao, James C},
  journal={Journal of Experimental Biology},
  volume={207},
  number={20},
  pages={3495--3506},
  year={2004},
  publisher={Company of Biologists}
}

@article{peng2018collective,
  title={Collective locomotion of two closely spaced self-propelled flapping plates},
  author={Peng, Ze-Rui and Huang, Haibo and Lu, Xi-Yun},
  journal={Journal of Fluid Mechanics},
  volume={849},
  pages={1068--1095},
  year={2018},
  publisher={Cambridge University Press}
}

@article{alben2021collective,
  title={Collective locomotion of two-dimensional lattices of flapping plates. Part 2. Lattice flows and propulsive efficiency},
  author={Alben, Silas},
  journal={Journal of Fluid Mechanics},
  volume={915},
  pages={A21},
  year={2021},
  publisher={Cambridge University Press}
}

@article{lin2022two,
  title={Two-dimensional hydrodynamic schooling of two flapping swimmers initially in tandem formation},
  author={Lin, Xingjian and Wu, Jie and Yang, Liming and Dong, Hao},
  journal={Journal of Fluid Mechanics},
  volume={941},
  pages={A29},
  year={2022},
  publisher={Cambridge University Press}
}

@article{lin2021flow,
  title={Flow-mediated organization of two freely flapping swimmers},
  author={Lin, Xingjian and Wu, Jie and Zhang, Tongwei and Yang, Liming},
  journal={Journal of Fluid Mechanics},
  volume={912},
  pages={A37},
  year={2021},
  publisher={Cambridge University Press}
}

@article{kelly2024effects,
  title={Effects of body shape on hydrodynamic interactions in a dense diamond fish school},
  author={Kelly, John and Dong, Haibo},
  journal={Physics of Fluids},
  volume={36},
  number={3},
  year={2024},
  publisher={AIP Publishing}
}

@article{pan2022effects,
  title={Effects of phase difference on hydrodynamic interactions and wake patterns in high-density fish schools},
  author={Pan, Yu and Dong, Haibo},
  journal={Physics of Fluids},
  volume={34},
  number={11},
  year={2022},
  publisher={AIP Publishing}
}

@article{weihs1975some,
  title={Some hydrodynamical aspects of fish schooling},
  author={Weihs, Daniel},
  journal={Swimming and Flying in Nature: Volume 2},
  pages={703--718},
  year={1975},
  publisher={Springer}
}

@article{zhang2024energy,
  title={Energy conservation by collective movement in schooling fish},
  author={Zhang, Yangfan and Lauder, George V},
  journal={Elife},
  volume={12},
  pages={RP90352},
  year={2024},
  publisher={eLife Sciences Publications Limited}
}

@article{kurt2021high,
  title={High-efficiency can be achieved for non-uniformly flexible pitching hydrofoils via tailored collective interactions},
  author={Kurt, Melike and Mivehchi, Amin and Moored, Keith},
  journal={Fluids},
  volume={6},
  number={7},
  pages={233},
  year={2021},
  publisher={MDPI}
}

@article{white2021tunabot,
  title={Tunabot Flex: A tuna-inspired robot with body flexibility improves high-performance swimming},
  author={White, Carl H and Lauder, George V and Bart-Smith, Hilary},
  journal={Bioinspiration \& Biomimetics},
  volume={16},
  number={2},
  pages={026019},
  year={2021},
  publisher={IOP Publishing}
}

@article{Gopalkrishnan1994,
  title={Active vorticity control in a shear flow using a flapping foil},
  author={Gopalkrishnan, R and Triantafyllou, Michael S and Triantafyllou, George S and Barrett, D},
  journal={Journal of Fluid Mechanics},
  volume={274},
  pages={1--21},
  year={1994},
  publisher={Cambridge University Press}
}

@article{Akhtar2007,
  title={Hydrodynamics of a biologically inspired tandem flapping foil configuration},
  author={Akhtar, Imran and Mittal, Rajat and Lauder, George V and Drucker, Elliot},
  journal={Theoretical and Computational Fluid Dynamics},
  volume={21},
  number={3},
  pages={155--170},
  year={2007},
  publisher={Springer}
}

@article{Quinn2014,
  title={Unsteady propulsion near a solid boundary},
  author={Quinn, Daniel B and Moored, Keith W and Dewey, Peter A and Smits, Alexander J},
  journal={Journal of Fluid Mechanics},
  volume={742},
  pages={152--170},
  year={2014},
  publisher={Cambridge University Press}
}

@article{Daghooghi2015,
  title={The hydrodynamic advantages of synchronized swimming in a rectangular pattern},
  author={Daghooghi, Mohsen and Borazjani, Iman},
  journal={Bioinspiration \& biomimetics},
  volume={10},
  number={5},
  pages={056018},
  year={2015},
  publisher={IOP Publishing}
}

@misc{han2023revealing,
      title={Revealing the mechanism and scaling laws behind equilibrium altitudes of near-ground pitching hydrofoils}, 
      author={Tianjun Han and Qiang Zhong and Amin Mivehchi and Daniel B. Quinn and Keith W. Moored},
      year={2023},
      eprint={2304.14562},
      archivePrefix={arXiv},
      primaryClass={physics.flu-dyn}
}

@article{Dong2007,
author = {Dong, Gen Jin and Lu, Xi Yun},
doi = {10.1063/1.2736083},
issn = {10706631},
journal = {Physics of Fluids},
number = {5},
title = {{Characteristics of flow over traveling wavy foils in a side-by-side arrangement}},
volume = {19},
year = {2007}
}

@article{Heydari2020,
archivePrefix = {arXiv},
arxivId = {2009.12715},
author = {Heydari, Sina and Kanso, Eva},
eprint = {2009.12715},
journal = {arXiv},
number = {12715v1},
pages = {1--16},
title = {{School cohesion, speed, and efficiency are modulated by the swimmers flapping motion}},
url = {http://arxiv.org/abs/2009.12715},
volume = {2009:},
year = {2020}
}

@article{Dewey2014,
author = {Dewey, P. A. and Quinn, D. B. and Boschitsch, B. M. and Smits, A. J.},
doi = {10.1063/1.4871024},
issn = {1070-6631},
journal = {Physics of Fluids},
month = {apr},
number = {4},
pages = {041903},
title = {{Propulsive performance of unsteady tandem hydrofoils in a side-by-side configuration}},
url = {http://scitation.aip.org/content/aip/journal/pof2/26/4/10.1063/1.4871024},
volume = {26},
year = {2014}
}

@article{Boschitsch2014,
author = {Boschitsch, Birgitt M. and Dewey, Peter a. and Smits, Alexander J.},
doi = {10.1063/1.4872308},
issn = {1070-6631},
journal = {Physics of Fluids},
month = {may},
number = {5},
pages = {051901},
title = {{Propulsive performance of unsteady tandem hydrofoils in an in-line configuration}},
url = {http://scitation.aip.org/content/aip/journal/pof2/26/5/10.1063/1.4872308},
volume = {26},
year = {2014}
}

@article{Kurt2020,
author = {Kurt, M. and Panah, A. E. and Moored, K. W.},
doi = {10.3390/biomimetics5020013},
issn = {2313-7673},
journal = {Biomimetics},
number = {2},
pages = {13},
title = {{Flow interactions between low aspect ratio hydrofoils in In-line and staggered arrangements}},
volume = {5},
year = {2020}
}

@article{Ramananarivo2016,
author = {Ramananarivo, Sophie and Fang, Fang and Oza, Anand and Zhang, Jun and Ristroph, Leif},
doi = {10.1103/PhysRevFluids.1.071201},
issn = {2469990X},
journal = {Physical Review Fluids},
number = {7},
pages = {1--9},
title = {{Flow interactions lead to orderly formations of flapping wings in forward flight}},
volume = {1},
year = {2016}
}

@article{kurt2018flow,
  title={Flow interactions of two-and three-dimensional networked bio-inspired control elements in an in-line arrangement},
  author={Kurt, Melike and Moored, Keith W},
  journal={Bioinspiration \& biomimetics},
  volume={13},
  number={4},
  pages={045002},
  year={2018},
  publisher={IOP Publishing}
}

@article{newbolt2019flow,
  title={Flow interactions between uncoordinated flapping swimmers give rise to group cohesion},
  author={Newbolt, Joel W and Zhang, Jun and Ristroph, Leif},
  journal={Proceedings of the National Academy of Sciences},
  volume={116},
  number={7},
  pages={2419--2424},
  year={2019},
  publisher={National Acad Sciences}
}

@article{moored2019inviscid,
  title={Inviscid scaling laws of a self-propelled pitching airfoil},
  author={Moored, Keith W and Quinn, Daniel B},
  journal={AIAA Journal},
  volume={57},
  number={9},
  pages={3686--3700},
  year={2019},
  publisher={American Institute of Aeronautics and Astronautics}
}

@article{weihs1973hydromechanics,
  title={Hydromechanics of fish schooling},
  author={Weihs, D},
  journal={Nature},
  volume={241},
  number={5387},
  pages={290--291},
  year={1973},
  publisher={Nature Publishing Group}
}

@article{portugal2014upwash,
  title={Upwash exploitation and downwash avoidance by flap phasing in ibis formation flight},
  author={Portugal, Steven J and Hubel, Tatjana Y and Fritz, Johannes and Heese, Stefanie and Trobe, Daniela and Voelkl, Bernhard and Hailes, Stephen and Wilson, Alan M and Usherwood, James R},
  journal={Nature},
  volume={505},
  number={7483},
  pages={399--402},
  year={2014},
  publisher={Nature Research}
}

@article{muscutt2017performance,
  title={Performance augmentation mechanism of in-line tandem flapping foils},
  author={Muscutt, LE and Weymouth, GD and Ganapathisubramani, Bharathram},
  journal={Journal of Fluid Mechanics},
  volume={827},
  pages={484--505},
  year={2017},
  publisher={Cambridge University Press}
}

@article{becker2015hydrodynamic,
  title={Hydrodynamic schooling of flapping swimmers},
  author={Becker, Alexander D and Masoud, Hassan and Newbolt, Joel W and Shelley, Michael and Ristroph, Leif},
  journal={Nature communications},
  volume={6},
  number={1},
  pages={1--8},
  year={2015},
  publisher={Nature Publishing Group}
}

@article{verma2018efficient,
author = {Verma, Siddhartha and Novati, Guido and Koumoutsakos, Petros and Sethian, James A},
journal = {Proceedings of the National Academy of Sciences},
number = {23},
pages = {5849--5854},
title = {{Efficient collective swimming by harnessing vortices through deep reinforcement learning}},
volume = {115},
year = {2018}
}

@article{ramananarivo2016flow,
  title={Flow interactions lead to orderly formations of flapping wings in forward flight},
  author={Ramananarivo, Sophie and Fang, Fang and Oza, Anand and Zhang, Jun and Ristroph, Leif},
  journal={Physical Review Fluids},
  volume={1},
  number={7},
  pages={071201},
  year={2016},
  publisher={APS}
}

@article{li2019energetics,
  title={On the energetics and stability of a minimal fish school},
  author={Li, Gen and Kolomenskiy, Dmitry and Liu, Hao and Thiria, Benjamin and Godoy-Diana, Ramiro},
  journal={PLoS One},
  volume={14},
  number={8},
  pages={e0215265},
  year={2019},
  publisher={Public Library of Science San Francisco, CA USA}
}

@article{smits2019undulatory,
  title={Undulatory and oscillatory swimming},
  author={Smits, Alexander J},
  journal={Journal of Fluid Mechanics},
  volume={874},
  pages={P1},
  year={2019},
  publisher={Cambridge University Press}
}

@article{Lee2023,
author = {Lee, David S. and Hrynuk, John T. and Moored, Keith W.},
title = {Effects of Spanwise Spacing on the Interaction of Tandem Pitching Hydrofoils},
journal = {AIAA Journal},
volume = {61},
number = {11},
pages = {5121-5131},
year = {2023},
doi = {10.2514/1.J063077}
}

@article{berlinger2021implicit,
  title={Implicit coordination for 3D underwater collective behaviors in a fish-inspired robot swarm},
  author={Berlinger, Florian and Gauci, Melvin and Nagpal, Radhika},
  journal={Science Robotics},
  volume={6},
  number={50},
  pages={eabd8668},
  year={2021},
  publisher={American Association for the Advancement of Science}
}

@article{katzschmann2018exploration,
  title={Exploration of underwater life with an acoustically controlled soft robotic fish},
  author={Katzschmann, Robert K and DelPreto, Joseph and MacCurdy, Robert and Rus, Daniela},
  journal={Science Robotics},
  volume={3},
  number={16},
  pages={eaar3449},
  year={2018},
  publisher={American Association for the Advancement of Science}
}

@article{zhu2019tuna,
  title={Tuna robotics: A high-frequency experimental platform exploring the performance space of swimming fishes},
  author={Zhu, Joseph and White, Carl and Wainwright, Dylan K and Di Santo, Valentina and Lauder, George V and Bart-Smith, Hilary},
  journal={Science Robotics},
  volume={4},
  number={34},
  pages={eaax4615},
  year={2019},
  publisher={American Association for the Advancement of Science}
}

@article{zhong2021tunable,
  title={Tunable stiffness enables fast and efficient swimming in fish-like robots},
  author={Zhong, Qiang and Zhu, Joseph and Fish, Frank E and Kerr, Steven John and Downs, AM and Bart-Smith, Hilary and Quinn, DB},
  journal={Science Robotics},
  volume={6},
  number={57},
  pages={eabe4088},
  year={2021},
  publisher={American Association for the Advancement of Science}
}

@article{zhong2021aspect,
  title={Aspect ratio affects the equilibrium altitude of near-ground swimmers},
  author={Zhong, Qiang and Han, Tianjun and Moored, Keith W and Quinn, Daniel B},
  journal={Journal of Fluid Mechanics},
  volume={917},
  pages={A36},
  year={2021},
  publisher={Cambridge University Press}
}

@article{roper2011review,
  title={A review of developments towards biologically inspired propulsion systems for autonomous underwater vehicles},
  author={Roper, DT and Sharma, S and Sutton, R and Culverhouse, P},
  journal={Proceedings of the Institution of Mechanical Engineers, Part M: Journal of Engineering for the Maritime Environment},
  volume={225},
  number={2},
  pages={77--96},
  year={2011},
  publisher={SAGE Publications Sage UK: London, England}
}

@article{lauder2011bioinspiration,
  title={Bioinspiration from fish for smart material design and function},
  author={Lauder, GV and Madden, PGA and Tangorra, JL and Anderson, E and Baker, TV},
  journal={Smart Materials and Structures},
  volume={20},
  number={9},
  pages={094014},
  year={2011},
  publisher={IOP Publishing}
}

@article{siddall2014launching,
  title={Launching the AquaMAV: bioinspired design for aerial--aquatic robotic platforms},
  author={Siddall, R and Kova{\v{c}}, M},
  journal={Bioinspiration \& biomimetics},
  volume={9},
  number={3},
  pages={031001},
  year={2014},
  publisher={IOP Publishing}
}

@article{fish2020bio,
  title={Bio-inspired aquatic drones: Overview},
  author={Fish, Frank E},
  journal={Bioinspiration \& Biomimetics},
  pages={060401},
  year={2020},
  publisher={IOP Publishing}
}

@article{king2018experimental,
  title={Experimental observations of the three-dimensional wake structures and dynamics generated by a rigid, bioinspired pitching panel},
  author={King, Justin T and Kumar, Rajeev and Green, Melissa A},
  journal={Physical Review Fluids},
  volume={3},
  number={3},
  pages={034701},
  year={2018},
  publisher={APS}
}

@article{zhu2023flow,
  title={Flow-induced oscillations of pitching swept wings: stability boundary, vortex dynamics and force partitioning},
  author={Zhu, Yuanhang and Breuer, Kenneth},
  journal={Journal of Fluid Mechanics},
  volume={977},
  pages={A1},
  year={2023},
  publisher={Cambridge University Press}
}

@article{jeong1995identification,
  title={On the identification of a vortex},
  author={Jeong, Jinhee and Hussain, Fazle},
  journal={Journal of fluid mechanics},
  volume={285},
  pages={69--94},
  year={1995},
  publisher={Cambridge University Press}
}

@article{ormonde2024two,
  title={Two-dimensionally stable self-organisation arises in simple schooling swimmers through hydrodynamic interactions},
  author={Ormonde, Pedro C and Kurt, Melike and Mivehchi, Amin and Moored, Keith W},
  journal={Journal of Fluid Mechanics},
  volume={1000},
  pages={A90},
  year={2024},
  publisher={Cambridge University Press}
}

@article{buchholz2008wake,
  title={The wake structure and thrust performance of a rigid low-aspect-ratio pitching panel},
  author={Buchholz, James HJ and Smits, Alexander J},
  journal={Journal of fluid mechanics},
  volume={603},
  pages={331--365},
  year={2008},
  publisher={Cambridge University Press}
}

@article{ormonde2021two,
  title={Two-dimensionally stable self-organization arises in simple schooling swimmers through hydrodynamic interactions},
  author={Ormonde, Pedro Costa and Kurt, Melike and Mivehchi, Amin and Moored, Keith W},
  journal={arXiv preprint arXiv:2102.03571},
  year={2021}
}

@article{ligman2023comprehensive,
  title={A comprehensive review of hydrodynamic studies on fish schooling},
  author={Ligman, Montana Genivieve and Lund, Joshua and F{\"u}rth, Mirjam},
  journal={Bioinspiration \& Biomimetics},
  year={2023}
}

@article{liao2003karman,
  title={The K{\'a}rm{\'a}n gait: novel body kinematics of rainbow trout swimming in a vortex street},
  author={Liao, James C and Beal, David N and Lauder, George V and Triantafyllou, Michael S},
  journal={Journal of experimental biology},
  volume={206},
  number={6},
  pages={1059--1073},
  year={2003},
  publisher={Company of Biologists}
}

@article{hemelrijk2015increased,
  title={The increased efficiency of fish swimming in a school},
  author={Hemelrijk, Ch K and Reid, DAP and Hildenbrandt, H and Padding, JT},
  journal={Fish and Fisheries},
  volume={16},
  number={3},
  pages={511--521},
  year={2015},
  publisher={Wiley Online Library}
}

@article{stocker1999models,
  title={Models for tuna school formation},
  author={St{\"o}cker, Sabine},
  journal={Mathematical biosciences},
  volume={156},
  number={1-2},
  pages={167--190},
  year={1999},
  publisher={Elsevier}
}

@article{wei2023hydrodynamic,
  title={Hydrodynamic interactions and wake dynamics of fish schooling in rectangle and diamond formations},
  author={Wei, Chang and Hu, Qiao and Li, Shijie and Shi, Xindong},
  journal={Ocean Engineering},
  volume={267},
  pages={113258},
  year={2023},
  publisher={Elsevier}
}

@article{kelly2023hydrodynamics,
  title={Hydrodynamics of body--body interactions in dense synchronous elongated fish schools},
  author={Kelly, John and Pan, Yu and Menzer, Alec and Dong, Haibo},
  journal={Physics of Fluids},
  volume={35},
  number={4},
  year={2023},
  publisher={AIP Publishing}
}

@article{li2020vortex,
  title={Vortex phase matching as a strategy for schooling in robots and in fish},
  author={Li, Liang and Nagy, M{\'a}t{\'e} and Graving, Jacob M and Bak-Coleman, Joseph and Xie, Guangming and Couzin, Iain D},
  journal={Nature communications},
  volume={11},
  number={1},
  pages={5408},
  year={2020},
  publisher={Nature Publishing Group UK London}
}

@article{heydari2021school,
  title={School cohesion, speed and efficiency are modulated by the swimmers flapping motion},
  author={Heydari, Sina and Kanso, Eva},
  journal={Journal of Fluid Mechanics},
  volume={922},
  pages={A27},
  year={2021},
  publisher={Cambridge University Press}
}

@article{ribeiro2021wake,
  title={Wake-foil interactions and energy harvesting efficiency in tandem oscillating foils},
  author={Ribeiro, Bernardo Luiz R and Su, Yunxing and Guillaumin, Quentin and Breuer, Kenneth S and Franck, Jennifer A},
  journal={Physical Review Fluids},
  volume={6},
  number={7},
  pages={074703},
  year={2021},
  publisher={APS}
}

@article{zhu2025wavenumber,
  title={Wavenumber affects the lift of ray-inspired fins near a substrate},
  author={Zhu, Yuanhang and Liu, Leo and Han, Tianjun and Feng, Qimin and Moored, Keith W and Zhong, Qiang and Quinn, Daniel B},
  journal={Journal of the Royal Society Interface},
  volume = {22},
  number = {231},
  pages = {20250276},
  year = {2025},
  month = {10},
  publisher={The Royal Society}
}

\end{document}